\documentclass[12pt,draftcls,journal,onecolumn]{IEEEtran}
\usepackage{xspace}
\usepackage{amsmath}
\usepackage{amssymb}
\usepackage{amscd}
\usepackage{xypic}
\xyoption{curve}
\usepackage{latexsym}
\usepackage{theorem}
\usepackage{paralist} % compact and in-paragraph lists
\usepackage[dvips]{graphicx}
\usepackage{epsfig}
\usepackage{algorithm}
\usepackage{algorithmic}
\usepackage{times}
\usepackage{floatflt} % text flows around figures
\newtheorem{lemma}{Lemma}
\newtheorem{theorem}{Theorem}
\newtheorem{corollary}{Corollary}

\begin{document}
\title{Linear Time Encoding of LDPC Codes}
\author{Jin~Lu
        and~Jos\'e~M.~F.~Moura,~\IEEEmembership{Fellow,~IEEE}% <-this % stops a space
\thanks{}
\thanks{Jin Lu}
\thanks{Sun Microsystems}
\thanks{1, StorageTek Drive, Louisville, CO 80026}
\thanks{Ph: +(303) 661-6961 Fax:+(303) 673-2431}
\thanks{Email: j.lu@sun.com}
\thanks{Jos\'e~M.~F.~Moura}
\thanks{Department of Electrical and Computer Engineering}
\thanks{Carnegie Mellon University, Pittsburgh, PA 15213}
\thanks{Ph: +(412) 268-6341 Fax:+(412) 268-3890}
\thanks{Email: moura@ece.cmu.edu}}
% <-this % stops a space
%\thanks{Jos\'e~M.~F.~Moura is with Carnegie Mellon University.}}
% note the % following the last \IEEEmembership and also the first \thanks -
% instance, "A\textbf{} \textbf{}B" will typeset as "A B" not "AB". If you want
% "AB" then you have to do: "A\textbf{}\textbf{}B"
\markboth{}{Shell \MakeLowercase{\textit{et al.}}: Bare Demo of
IEEEtran.cls for Journals}
% *** Note that you probably will NOT want to include the author's name in ***
% *** the headers of peer review papers.                                   ***

% If you want to put a publisher's ID mark on the page
% (can leave text blank if you just want to see how the
% text height on the first page will be reduced by IEEE)
%\pubid{0000--0000/00\$00.00~\copyright~2002 IEEE}

%\specialpapernotice{(Invited Paper)}

\maketitle
\begin{abstract}
In this paper, we propose a linear complexity encoding method for
arbitrary LDPC codes. We start from a simple graph-based encoding
method ``label-and-decide.'' We prove that the
``label-and-decide'' method is applicable to Tanner graphs with a
hierarchical structure---pseudo-trees--- and that the resulting
encoding complexity is linear with the code block length. Next, we
define a second type of Tanner graphs---the encoding stopping set.
The encoding stopping set is encoded in linear complexity by a
revised label-and-decide algorithm---the
''label-decide-recompute.'' Finally, we prove that any Tanner
graph can be partitioned into encoding stopping sets and
pseudo-trees. By encoding each encoding stopping set or
pseudo-tree sequentially, we develop a linear complexity encoding
method for general LDPC codes where the encoding complexity is
proved to be less than~$4 \cdot M \cdot (\overline{k} - 1)$, where
$M$ is the number of independent rows in the parity check matrix
and $\overline{k}$ represents the mean row weight of the parity
check matrix.
\end{abstract}


\IEEEpeerreviewmaketitle

\newpage
\section{Introduction}
\label{sec:intro}
 Low Density Parity Check (LDPC) codes~\cite{Gallager} are excellent error
correcting codes with performance close to the Shannon
Capacity~\cite{Mackay}. The key weakness of LDPC codes is their
apparently high encoding complexity. The conventional way to
encode LDPC codes is to multiply the data words
$\overrightarrow{s}$ by the code generator matrix~$\mathbf{G}$,
i.e., the code words are $\overrightarrow{x} = \mathbf{G} \cdot
\overrightarrow{s}$. Though the parity-check matrix~$\mathbf{H}$
for LDPC codes is sparse, the associated generator
matrix~$\mathbf{G}$ is not. The encoding complexity of LDPC codes
is~$\mathcal{O}(n^2)$ where $n$ is the block length of the LDPC
code. For moderate to high code block length~$n$, this quadratic
behavior is very significant and it severely affects the
application of LDPC codes. For example, LDPC codes have advantages
over turbo codes~\cite{Berrou} in almost every aspect except that
LDPC codes have $\mathcal{O}(n^2)$~encoding complexity, while
turbo codes have $\mathcal{O}(n)$~encoding complexity. It is
highly desirable to reduce the $\mathcal{O}(n^2)$~encoding
complexity of LDPC codes.

Several authors have addressed the issue of speeding encoding of
LDPC codes and, largely speaking, they follow three different
paths. The first path designs efficient encoding methods for
particular types of LDPC codes. We list a few typical
representers. Reference~\cite{Jin:1} proposes a linear complexity
encoding method for cycle codes---LDPC codes with column weight~2.
Reference~\cite{Li} presents an efficient encoder for quasi-cyclic
LDPC codes. In~\cite{Mittelholzer}, an efficient encoding approach
is proposed for Reed-Solomon-type array codes.
Reference~\cite{Jin:2} shows that there exists a linear time
encoder for turbo-structured LDPC codes. Reference~\cite{Kou}
constructs LDPC codes based on finite geometries and proves that
this type of structured LDPC codes can be encoded in linear time.
In~\cite{Johnson,Ryan}, two families of irregular LDPC Codes with
cyclic structure and low encoding complexity are designed. In
addition, an approximately lower triangular ensemble of LDPC
Codes~\cite{Burshtein} was proposed to facilitate almost linear
complexity encoding. The above low complexity encoders are only
applicable to a small subset of LDPC codes, and some of the LDPC
codes discussed above have performance loss when compared to
randomly constructed LDPC codes. The second path borrows the
decoder architecture and encodes LDPC codes iteratively on their
Tanner graphs~\cite{David:1,David:2}. The iterative LDPC encoding
algorithm is easy to implement. However, there is no guarantee
that iterative encoding will successfully get the codeword. In
particular, the iterative encoding method will get trapped at the
stopping set. The third path utilizes the sparseness of the parity
check matrix to design a low complexity encoder.
In~\cite{Richardson}, the authors present an algorithm named
``greedy search'' that reduces the coefficient of the quadratic
term. This encoding method is relatively efficient. Its
computation complexity and matrix storage need to be further
reduced for most practical applications.

In this paper, we develop an \emph{exact} linear complexity
encoding method for arbitrary LDPC codes. We start from two
particular Tanner graph structures---``pseudo-tree'' and
``encoding stopping set''--- and prove that both the pseudo-tree
and the encoding stopping set LDPC codes can be encoded in linear
time. Next, we prove that any LDPC code with maximum column weight
three can be decomposed into pseudo-trees and encoding stopping
sets. Therefore, LDPC codes with maximum column weight three can
be encoded in linear time and the encoding complexity is no more
than~$2 \cdot M \cdot (\overline{k} - 1)$ where $M$ denotes the
number of independent rows of the parity check matrix and
$\overline{k}$ represents the average row weight. Finally, we
extend the $\mathcal{O}(n)$~complexity encoder to LDPC codes with
arbitrary row weight distributions and column weight
distributions. For arbitrary LDPC codes, we achieve
$\mathcal{O}(n)$ encoding complexity, not exceeding~$4 \cdot M
\cdot (\overline{k} - 1)$.

The remaining of the paper is organized as follows. In
Section~\ref{sec:notation}, we introduce relevant definitions and
notation. Section~\ref{sec:label-and-decide} proposes a simple
encoding algorithm ``label-and-decide'' that directly encodes an
LDPC code on its Tanner graph. Section~\ref{sec:pseudotree}
presents a particular type of Tanner graph with
multi-layers---``pseudo-tree'' and proves that any pseudo-tree can
be encoded successfully in linear time by the label-and-decide
algorithm. Section~\ref{sec:encoding stopping set} studies the
complement of the pseudo-tree---``encoding stopping set.''
Section~\ref{sec:linear-encoding-stoppingset} proves that the
encoding stopping set can also be encoded in linear time by an
encoding method named ``label-decide-recompute.''
Section~\ref{sec:linear-encoding-ldpccodes} demonstrates that any
LDPC code with column weight at most three can be decomposed into
pseudo-trees and encoding stopping sets. By encoding each
pseudo-tree or encoding stopping set sequentially using the
label-and-decide or the label-decide-recompute algorithms, we
achieve linear complexity encoding for LDPC codes with maximum
column weight three. Finally, we extend in this Section this
linear time encoding method to LDPC codes with arbitrary column
weight distributions and row weight distributions.
Section~\ref{sec:future} concludes the paper.

\section{Notation}
\label{sec:notation}
%\label{sec:intro}
{\bf LDPC codes.} LDPC codes can be described by their
parity-check matrix or their
 associated Tanner graph~\cite{Tanner}. In the Tanner graph, each bit
becomes a bit node and each parity-check constraint becomes a
check node. If a bit is involved in a parity-check constraint,
there is an edge connecting the  bit node and the corresponding
check node. The degree of a check node in a Tanner graph is
equivalent to the number of one's in the corresponding row of the
parity check matrix, or, in another words, the row weight of the
corresponding row. We will use the term ``degree of a check node''
and ``row weight'' interchangeably in this paper. Similarly, the
degree of a bit node in a Tanner graph is equivalent to the column
weight of the corresponding column of the parity check matrix, and
we will interchangeably use the term ``degree of a bit node'' and
``column weight'' in this paper. The LDPC codes discussed in this
paper may be irregular, i.e., different columns of the parity
check matrix have different column weights and different rows of
the parity check matrix have different row weights. The parity
check matrix of an LDPC code may not be of full rank. If a row in
the parity check matrix can be written as the binary sums of some
other rows in the parity check matrix, this row is said to be
\emph{dependent} on the other rows. Otherwise, it is an
\emph{independent} row.

{\bf Arithmetic over the binary field.}  We represent by
``$\oplus$'' the summation over the binary field, i.e., an XOR
operation. For example, $0 \oplus 1 = \mod(0+1\,,\,2) = 1$.
Similarly, we have $0 \oplus 0 = 0$, $1 \oplus 0 = 1$, $1 \oplus 1
= 0$. In addition, we have the following equation $-x = x$ in the
binary field.

{\bf Generalized parity check equation.} A conventional parity
check equation is shown in~(\ref{equ11}). The right-hand side of
the parity check equation is always 0.
\begin{equation}
\label{equ11} x_1 \oplus x_2 \oplus \ldots \oplus x_k = 0
\end{equation}
In this paper, we define the \emph{generalized parity check
equation}, as shown in~(\ref{equ12})
\begin{equation}
\label{equ12} x_1 \oplus x_2 \oplus \ldots \oplus x_k = b
\end{equation}
On the right-hand side of equation~(\ref{equ12}), $b$ is a
constant that can be either 0 or 1.

Let $C$ be a standard parity check equation. If the values of some
of the bits in the left-hand side of $C$ are already known, then
$C$ can be equivalently rewritten as a generalized parity check
equation. For example, if the values of the bits $x_{p+1},
x_{p+2}, \ldots, x_k$ are known, we move these bits from the
left-hand side of equation~(\ref{equ11}) to its right-hand side
and rewrite it as follows.
\begin{equation}
\label{equ13} x_1 \oplus x_2 \oplus \ldots \oplus x_p = x_{p+1}
\oplus x_{p+2} \oplus \ldots \oplus x_k = b
\end{equation}
Let $C_1$, $C_2$, $\ldots$, $C_p$ be $p$ generalized parity check
equations, as shown in~(\ref{equ14})
\begin{equation}
\label{equ14} \begin{array}{c}
\begin{array}{ccccccccc}
               C_1: &  x_{1,1} & \oplus & x_{1,2} & \oplus & \ldots & x_{1,a_1} & = & b_1 \\
               C_2: &  x_{2,1} & \oplus & x_{2,2} & \oplus & \ldots & x_{2,a_2} & = & b_2 \\
                      \vdots &  & \vdots &  &  & \vdots &  & \vdots \\
               C_p: &  x_{p,1} & \oplus & x_{p,2} & \oplus & \ldots & x_{p,a_p} & = &
                  b_p
                  \end{array}    \end{array}
\end{equation}

We say $C_1$, $C_2$, $\ldots$, $C_p$ are dependent on each other
if the corresponding homogeneous equations in~(\ref{equ15}) are
dependent on each other.
\begin{equation}
\label{equ15} \begin{array}{c}
\begin{array}{ccccccccc}
                  x_{1,1} & \oplus & x_{1,2} & \oplus & \ldots & x_{1,a_1} & = & 0 \\
                  x_{2,1} & \oplus & x_{2,2} & \oplus & \ldots & x_{2,a_2} & = & 0 \\
                  \vdots &  & \vdots &  &  & \vdots &  & \vdots \\
                  x_{p,1} & \oplus & x_{p,2} & \oplus & \ldots &
                  x_{p,a_p} & = & 0
                  \end{array}    \end{array}
\end{equation}
From equations~(\ref{equ14}) and~(\ref{equ15}), we derive that
\begin{equation}
\label{equ16} b_1 \oplus b_2 \oplus \ldots \oplus b_p = 0
\end{equation}
when the $p$ generalized parity check equations $C_1$, $C_2$,
$\ldots$, $C_p$ are dependent on each other.

{\bf Connected graph.} A graph is \emph{connected} if there exists
a path from any vertex to any other vertices in the graph. If a
graph is not connected, we call it a \emph{disjoint} graph.

{\bf Relative complement of a subgraph~$\mathcal{S}$ in a Tanner
graph~$\mathcal{G}$.} Let $\mathcal{G}$ be a Tanner graph and
$\mathcal{S}$ be a subgraph of $\mathcal{G}$, i.e.,
$\mathcal{S}\subset\mathcal{G}$. We use the symbol $\mathcal{G}
\backslash \mathcal{S}$ to denote the subgraph that contains the
nodes and edges in~$\mathcal{G}$, but not in~$\mathcal{S}$. For
example, let $C_1$, $C_2$, $\ldots$, $C_k$ be $k$ check nodes in a
Tanner graph~$\mathcal{G}$. The subgraph $\mathcal{G} \backslash
\{ C_1, C_2, \ldots, C_k \}$ represents the remaining graph after
deleting check nodes $C_1$, $C_2$, $\ldots$, $C_k$ from
$\mathcal{G}$. Assume $\mathcal{G}_1$, $\mathcal{G}_2$, $\ldots$,
$\mathcal{G}_k$ are $k$ subgraphs in a Tanner graph $\mathcal{G}$.
The notation $\mathcal{G} \backslash \{\mathcal{G}_1 \cup
\mathcal{G}_2 \cup \ldots \cup \mathcal{G}_k\}$ represents the
subgraph where nodes and edges are in $\mathcal{G}$, but not in
$\mathcal{G}_i$, $1 \leq i \leq k$.

\section{``Label-and-decide'' Encoding Algorithm}
\label{sec:label-and-decide}
Initially, Tanner
graphs~\cite{Tanner} were developed to explain the decoding
process for LDPC codes; in fact, they can be used for the encoding
of LDPC codes as well~\cite{David:1}. To encode an LDPC code using
its Tanner graph, we identify information bits and parity bits
through a labeling process on the graph. After determining the
information bits and the parity bits, we start by assigning
numerical values to the bit nodes labeled as information bits and
then in a second step, calculate the missing values of the parity
bits sequentially. This encoding approach is named
\emph{label-and-decide}. It is described in
Algorithm~\ref{encoding-algorithm1}.
\begin{algorithm}
\caption{Label-and-decide algorithm \label{encoding-algorithm1}}
\begin{algorithmic}
\STATE {\bf Preprocessing (carry out only once):} \STATE  Label
every bit node either as information bit or parity bit on the
Tanner graph. \STATE {\bf Encoding:} \STATE  $Flag \leftarrow 0$;
\STATE Get the values of all the bits labeled as information bits;
\WHILE {there are parity bits undetermined}
       \IF {there exists one undetermined parity
bit~$x$ that can be uniquely computed from the values of the
information bits and the already determined parity bits}
           \STATE Compute the value of~$x$.
       \ELSE
           \STATE $Flag \leftarrow 1$, exit the while loop.
       \ENDIF
\ENDWHILE
\IF {$Flag=1$}
    \STATE Encoding is unsuccessful.
\ELSE
    \STATE Output the encoded codeword.
\ENDIF
\end{algorithmic}
\end{algorithm}

\begin{figure}[htb]
\centerline{\epsfig{figure=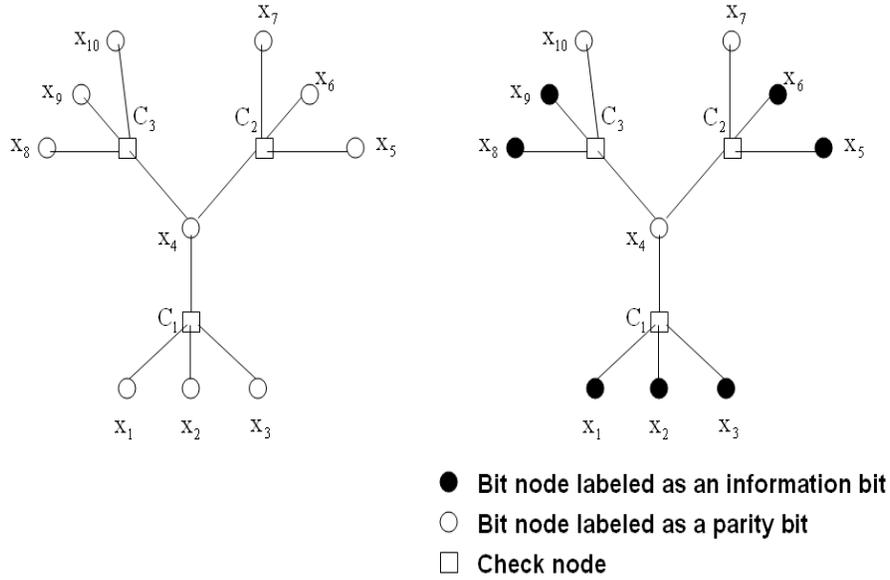,width=12cm,height=8cm}}
\caption{\label{encode1} Left: A Tanner graph. Right: Labeling bit
nodes on the Tanner graph shown on the left.}
\end{figure}
{\bf Example.} Figure~\ref{encode1} shows on the left an LDPC code
whose Tanner graph is a tree. Initially, all its bit nodes are
unlabeled. First, we randomly pick bit nodes~$x_{1}$, $x_{2}$, and
$x_{3}$ to be information bits. According to the parity check
equation~$C_1$, the value of bit~$x_{4}$ depends on the values of
the bits~$x_{1}$, $x_{2}$, and $x_{3}$ such that $x_{4} = x_{1}
\oplus x_{2} \oplus x_{3}$. Therefore, $x_{4}$ should be labeled
as a parity bit. Similarly, we label bits~$x_{5}$, $x_{6}$,
$x_{8}$, $x_{9}$ as information bits and label bits $x_{7}$,
$x_{10}$ as parity bits.
 We represent information
bits by solid circles and parity bits by empty circles. The
labeling result is shown on the right in Figure~\ref{encode1}.

By the above labeling process, we decide the systematic component
of the code word $\overrightarrow{x} = (x_{1} \,\, x_{2} \,\,
x_{3} \,\, x_{4} \,\, x_{5} \,\, x_{6} \,\, x_{7} \,\, x_{8} \,\,
x_{9} \,\, x_{10})$ to be $\overrightarrow{s} = (x_{1} \,\, x_{2}
\,\, x_{3} \,\, x_{5} \,\, x_{6} \,\, x_{8} \,\, x_{9})$ and the
parity component to be $\overrightarrow{p} = (x_{4} \,\, x_{7}
\,\, x_{10})$. The label-and-decide encoding on the code in
Figure~\ref{encode1} then has the following steps:
\begin{itemize}
\item[Step 1.] Get the values of the information bits~$x_{1}$,
$x_{2}$, $x_{3}$, $x_{5}$, $x_{6}$, $x_{8}$, and~$x_{9}$ from the
encoder input; \item[Step 2.] Compute the parity bit~$x_{4}$ from
the parity check equation~$C_1: x_{4}= x_{1} \oplus x_{2} \oplus
x_{3}$; \item[Step 3.] Compute the parity bit~$x_{7}$ from the
parity check equation~$C_2$: $x_{7}= x_{4} \oplus x_{5} \oplus
x_{6}$; compute the parity bit~$x_{10}$ from the parity check
equation~$C_3$: $x_{10}= x_{4} \oplus x_{8} \oplus x_{9}$.
\end{itemize}

In fact, any tree code (whose Tanner graph is cycle-free) can be
encoded in linear complexity by the label-and-decide algorithm. We
will prove this fact in Corollary~\ref{thm:corollary4} in
Section~\ref{sec:encoding stopping set}. Further, the
label-and-decide algorithm can be used to encode a particular type
of Tanner graphs with cycles, i.e., the pseudo-tree we propose in
the next section.

\section{Pseudo-tree}
\label{sec:pseudotree}  A \emph{pseudo-tree} is a connected Tanner
graph that satisfies the following conditions~(A1) through~(A4).
\begin{itemize}
\item[(A1)] It is composed of $2P + 1$~tiers where $P$ is a
positive integer. We number these tiers from~1 to~$2P + 1$,
starting from the top. The $(2i-1)$-th tier ($i = 1,2, \ldots
P+1$) contains only bit nodes, while the $(2i)$-th tier ($i = 1,2,
\ldots P$) contains only check nodes.
\item[(A2)] Each bit node in
the first tier has degree one and is connected to one and only one
check node in the second tier.
\item[(A3)] For each check
node~$C_\alpha$ in the $(2i)$-th tier, where $i$ can take any
value from~1 to~$P$, there is one and only one bit node~$x_\alpha$
in the $(2i-1)$-th tier (immediate upper tier) that connects
to~$C_\alpha$, and there are no other bit nodes in the upper tiers
that connect to~$C_\alpha$. We call $x_\alpha$ the \emph{parent}
of~$C_\alpha$ and $C_\alpha$ the \emph{child} of~$x_\alpha$.
\item[(A4)] For each bit node~$x_\beta$ in the $(2i-1)$-th tier,
where $i$ can take any value from~2 to~$P$, there is at most one
check node~$C_\beta$ in the $(2i)$-th tier (immediate lower tier)
that connects to~$x_\beta$, and there are no other check nodes in
the lower tiers that connect to~$x_\beta$.
\end{itemize}

\begin{figure}[htb]
\centerline{\epsfig{figure=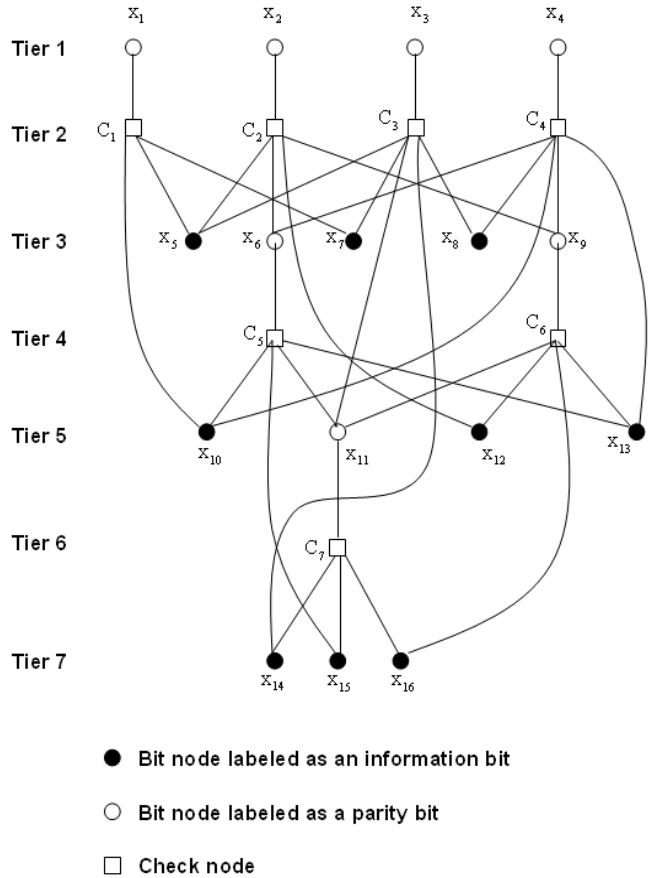,width=9cm,height=12cm}}
\caption{\label{encode2} A pseudo-tree.}
\end{figure}

For example, Figure~\ref{encode2} shows a pseudo-tree with seven
tiers. It contains many cycles. Each check node~$C_i$ in the
pseudo-tree is connected to a unique bit node in the immediate
upper tier, while each bit node~$x_i$ in the pseudo-tree may
connect to multiple check nodes in the upper tiers.

An important characteristic of a pseudo-tree is that it can be
encoded in linear complexity by the label-and-decide algorithm.
This is proved in the following lemma.
\begin{lemma}
\label{lemma1} Any LDPC code whose Tanner graph is a pseudo-tree
is linear time encodable.
\end{lemma}
\emph{Proof}: Let a pseudo-tree contain $2P + 1$~tiers, $N$~bit
nodes, and $M$~check nodes. Condition~(A3) guarantees that each
check node is connected to one and only one parent bit node in the
immediate upper tier. Condition~(A4) guarantees that different
check nodes are connected to different parent bit nodes.
 Therefore, there are
$M$~parent bit nodes for the check nodes. We label these
$M$~parent bit nodes as parity bits and the other $N-M$~bit nodes
as information bits.

The inputs of the encoder provide the values for all the
information bits. The task of the encoder is to compute the values
for all the parity bits. Let $x_\alpha$ be an arbitrary parity bit
in the $(2i-1)$-th tier. By conditions~(A3) and~(A4), there is
only one check node~$C_\alpha$ in the lower tiers that connects
to~$x_\alpha$. The value of~$x_\alpha$ is uniquely determined by
the parity check equation represented by~$C_\alpha$. According to
condition~(A3), all the bit nodes constrained by~$C_\alpha$ except
for~$x_\alpha$ are in tiers below the $(2i-1)$-th tier. Therefore,
the value of~$x_\alpha$ depends only on the values of the bit
nodes below the $(2i-1)$-th tier. For example, as shown in
Figure~\ref{encode2}, parity bit~$x_9$ is the parent bit node of
the check node~$C_6$. From the parity check equation~$C_6$, we see
that the value of~$x_9$ is computed from the values of~$x_{11}$,
$x_{12}$, $x_{16}$, and $x_{13}$, which are located below~$x_9$.
We compute the values of the parity bits tier by tier, starting
from the $(2P-1)$-th tier (bottom tier) and then progressing
upwards. Each time we compute the value of a parity bit, we only
need the values of those bits (both information bits and parity
bits) in lower tiers, which are already known. Hence, this
encoding process can proceed. The encoding process is repeated
until the values of all the parity bits in the first tier are
known.

We evaluate the computation complexity of the above encoding
process. Let $k_{i}$, $i=1, 2, \dots, M$, denote the number of
bits contained in the $i$-th~parity check equation. The
$i$-th~parity check equation determines the value of a parity bit
with $(k_{i}-2)$~XOR operations. So, $\sum_{i=1}^{M}
(k_{i}-2)$~XOR operations are required to obtain all the
$M$~parity bits. Let $\overline{k}=\frac{1}{M}\sum_{i=1}^{M}k_{i}$
denote the average number of bits in the $M$~parity check
equations, then the encoding complexity is~$\mathcal{O}(M
(\overline{k}-2))$. For LDPC codes with uniform row weight~$k$,
the encoding complexity is~$\mathcal{O}(M(k-2))$. The above
analysis shows that the encoding process is accomplished in linear
time. This completes the proof. \hfill $\Box$

We look at an example. We encode the pseudo-tree in
Figure~\ref{encode2} as follows:
\begin{itemize}
\item[Step 1.] Determine the values of all the information
bits~$x_{14}$, $x_{15}$, $x_{16}$, $x_{10}$, $x_{12}$, $x_{13}$,
$x_{5}$, $x_{7}$, and~$x_{8}$; \item[Step 2.] Compute the parity
bit~$x_{11}$ from the parity check equation~$C_7: x_{11}= x_{14}
\oplus x_{15} \oplus x_{16}$; \item[Step 3.] Compute the parity
bit~$x_{6}$ from the parity check equation~$C_5: x_{6}= x_{10}
\oplus x_{15} \oplus x_{11} \oplus x_{13}$; compute the parity
bit~$x_{9}$ from the parity check equation~$C_6: x_{9}= x_{11}
\oplus x_{12} \oplus x_{16} \oplus x_{13}$;
 \item[Step 4.] Compute the parity bits $x_1$, $x_2$, $x_3$,
and $x_4$ in the first tier by the parity check equations $C_1$,
$C_2$, $C_3$, and $C_4$ respectively: $x_1 = x_{10} \oplus x_{5}
\oplus x_{7}$, $x_2 = x_{5} \oplus x_{6} \oplus x_{12} \oplus
x_{9}$, $x_3 = x_{5} \oplus x_{7} \oplus x_{11} \oplus x_{14}
\oplus x_{8}$, $x_4 = x_{6} \oplus x_{8} \oplus x_{10} \oplus
x_{9} \oplus x_{13}$.
\end{itemize}
The above encoding process requires only 21 XOR operations.

\section{Encoding Stopping Set}
\label{sec:encoding stopping set} 
\begin{figure}[htb]
\centerline{\epsfig{figure=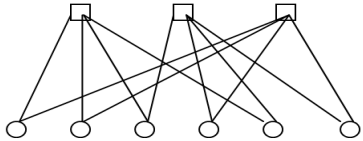,width=5cm,height=2cm}}
\caption{\label{encode3} A pseudo encoding stopping set.}
\end{figure}

\begin{figure}[htb]
\centerline{\epsfig{figure=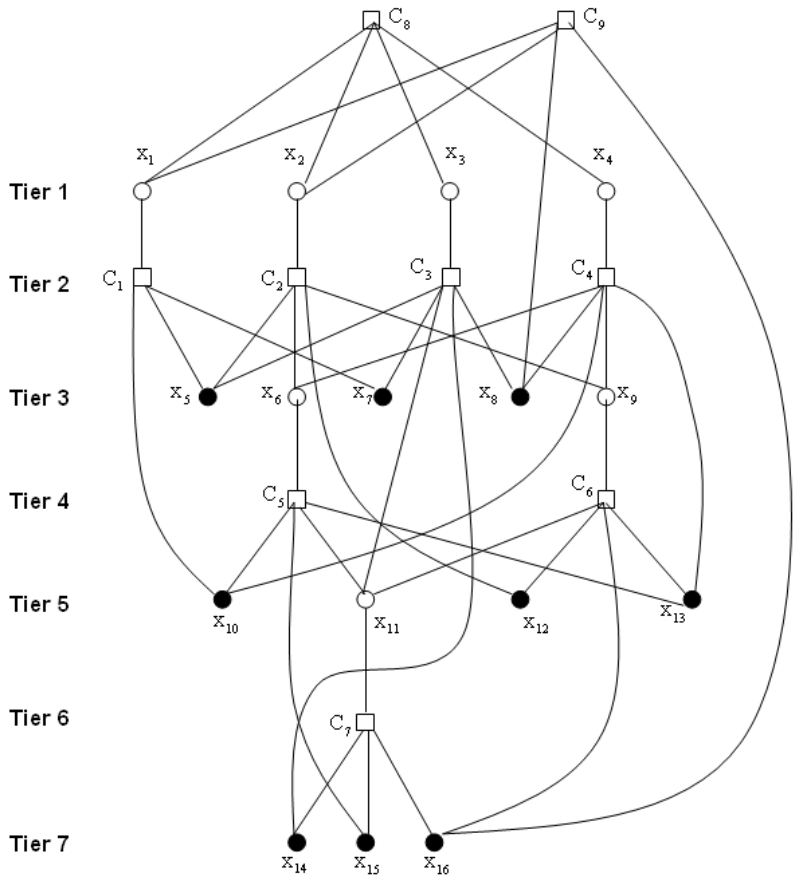,width=11cm,height=12cm}}
\caption{\label{encode4} An encoding stopping set whose proper
subgraph is a pseudo-tree shown in Figure~\ref{encode2}.}
\end{figure}

An \emph{encoding stopping set}
in a Tanner graph is a connected subgraph such that:
\begin{itemize}
\item[(B1)] If a check node~$C$ is in an encoding stopping set,
then all its associated bit nodes and the edges that are incident
on~$C$ are also in the encoding stopping set. \item[(B2)] Any bit
node in an encoding stopping set is connected to at least two
check nodes in the encoding stopping set. \item[(B3)] All the
check nodes included in an encoding stopping set are independent
of each other, i.e., any parity check equation can not be
represented as the binary sums of other parity check equations.
\end{itemize}
The number of check nodes in an encoding stopping set is called
its size. If a connected Tanner graph satisfies conditions (B1)
and (B2) but not condition (B3), we call this Tanner graph a
\emph{pseudo encoding stopping set}. For example, the Tanner graph
shown in Figure~\ref{encode3} is not an encoding stopping set but
a pseudo encoding stopping set since it satisfies conditions~(B1)
and~(B2) but not condition (B3).  The Tanner graph shown in
Figure~\ref{encode4} is an encoding stopping set. Its size is~9.
Every bit node in this encoding stopping set has degree greater
than or equal to~2, and every check node is independent of each
other. Please note that the ``encoding stopping set'' defined in
this paper is different from the ``stopping set'' defined
in~\cite{Di}. Stopping sets are used for the finite-length
analysis of LDPC codes on the binary erasure channel, while
encoding stopping sets are used here to develop efficient encoding
methods for LDPC codes.
From the above definitions of pseudo-tree and encoding stopping
set, we have the following lemma.
\begin{lemma}
\label{lemma9} Any pseudo-tree or union of pseudo-trees does not
contain encoding stopping sets.
\end{lemma}
The proof of Lemma~\ref{lemma9} is straightforward. We omit it
here.

We will show next that the label-and-decide algorithm can not
successfully encode encoding stopping sets.
\begin{theorem}
\label{thm:theorem1} An encoding stopping set can not be encoded
successfully by the label-and-decide algorithm.
\end{theorem}
\emph{Proof}: Let $\mathcal{E_s}$ be an encoding stopping set and
suppose $\mathcal{E_s}$ can be encoded successfully by the
label-and-decide algorithm. Let $x_\alpha$ be the last parity bit
being determined during the encoding process. Since
$\mathcal{E_s}$ is an encoding stopping set, $x_\alpha$ is
connected to at least two check nodes~$C_\beta$ and~$C_\gamma$ by
condition~(B2). Further, by condition~(B3), all the check nodes
in~$\mathcal{E_s}$, including~$C_\beta$ and~$C_\gamma$, are
independent of each other. Hence, for certain encoder inputs,
$C_\beta$ and~$C_\gamma$ provide different values for the parity
bit~$x_\alpha$. This contradicts the fact that every parity bit
can be uniquely determined successfully by the label-and-decide
algorithm. Hence, the label-and-decide algorithm can not encode an
encoding stopping set. This completes the proof. \hfill $\Box$

Conversely, if a Tanner graph does not contain any encoding
stopping set, there must exist a linear complexity encoder for the
corresponding code.
\begin{figure}[htb]
\centerline{\epsfig{figure=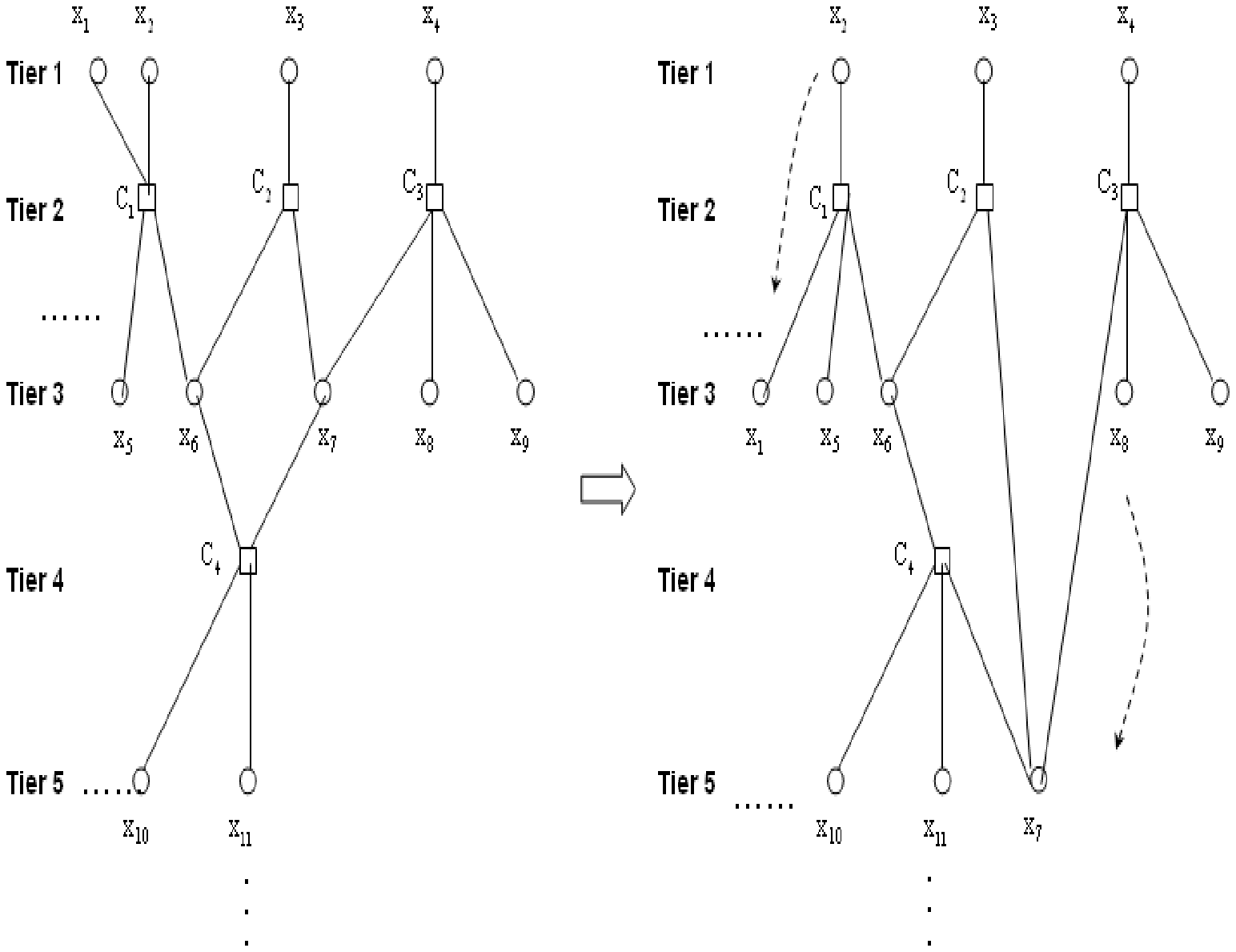,width=16cm,height=6cm}}
\caption{\label{encode5} Left: A multi-layer structure but not a
pseudo-tree. (Note that $C_4$ has two parents~$x_6$ and~$x_7$ and
$C_1$ has two parents~$x_1$ and~$x_2$.) Right: The pseudo-tree
that evolves from the multi-layer structure shown on the left.}
\end{figure}
\begin{theorem}
\label{thm:theorem2} If a Tanner graph~$\mathcal{G}$ does not
contain any encoding stopping set, then it can be encoded in
linear time by the label-and-decide algorithm.
\end{theorem}
\emph{Proof}:  We first delete all redundant check nodes (i.e.,
dependent on other check nodes) from the Tanner
graph~$\mathcal{G}$.  Next, we restrict our attention to the case
that $\mathcal{G}$ is a connected graph. We will show that
$\mathcal{G}$ can be equivalently transformed into a pseudo-tree
if it is free of any encoding stopping set. Since $\mathcal{G}$
does not contain any encoding stopping set, $\mathcal{G}$ itself
is not an encoding stopping set. Hence, there exist some
degree-one bit nodes in~$\mathcal{G}$. We generate a multi-layer
graph~$\mathcal{T}$ and place those degree-one bit nodes in the
first tier of~$\mathcal{T}$. Next, the check nodes that connect to
the degree-one bit nodes in the first tier of~$\mathcal{T}$ are
placed in the second tier of~$\mathcal{T}$. Notice that there
exist at least one bit node~$x_\alpha$ in~$\mathcal{G} \backslash
\mathcal{T}$ such that $x_\alpha$ connects to at most one check
node in~$\mathcal{G} \backslash \mathcal{T}$. This statement is
true. Otherwise, $\mathcal{G} \backslash \mathcal{T}$ becomes an
encoding stopping set, which contradicts the fact that
$\mathcal{G}$ does not contain any encoding stopping set. We pick
all the bit nodes in~$\mathcal{G} \backslash \mathcal{T}$ that
connect to at most one check node in~$\mathcal{G} \backslash
\mathcal{T}$ and place them in the third tier of $\mathcal{T}$.
Correspondingly, those check nodes in~$\mathcal{G} \backslash
\mathcal{T}$ that connect to the bit nodes in the third tier
of~$\mathcal{T}$
 are placed in the fourth tier of~$\mathcal{T}$.
Each time we find bit nodes in~$\mathcal{G} \backslash
\mathcal{T}$ that connect to at most one check node
in~$\mathcal{G} \backslash \mathcal{T}$, we place those bit nodes
in a new tier~$2s+1$ of~$\mathcal{T}$ and place the check nodes
connecting to those bit nodes in the following new tier~$2s+2$
of~$\mathcal{T}$. We continue finding such bit nodes and
increasing tiers till all the nodes in~$\mathcal{G}$ are included
in~$\mathcal{T}$. Up to now, the multi-layer structure constructed
so far satisfies the conditions~(A1), (A2), and~(A4). Condition
(A3) may fail to be satisfied. For example, as shown on the left
in Figure~\ref{encode5}, the check node~$C_4$ in tier~4 is
connected to two bit nodes~$x_6$ and~$x_7$ in tier~3, which
contradicts condition (A3). To satisfy condition (A3), we further
adjust the positions of the bit nodes. If a check node in
tier~$2i$ is connected to $k$~bit nodes in the upper tiers
of~$\mathcal{T}$, we pick one bit node in tier~$2i-1$ from these
$k$ bit nodes and leave its position unchanged. Next, we drag the
other $k-1$ bit nodes from their initial positions in tier~$2i-1$
to the ($2i+1$)-th tier. To illustrate, let us focus on
Figure~\ref{encode5} again. We drag the bit node~$x_7$ from tier~3
to tier~5 and drag the bit node~$x_1$ from tier~1 to tier~3. The
newly formed graph is shown on the right in Figure~\ref{encode5},
which follows condition~(A3). By tuning the positions of the bit
nodes in this way, the resulting hierarchical graph satisfies
conditions~(A1) to~(A4). In this way, we transform~$\mathcal{G}$
into a pseudo-tree. By Lemma~\ref{lemma1}, a pseudo-tree is linear
time encodable. Therefore, the encoding complexity
of~$\mathcal{G}$ is~$\mathcal{O}(M)$ where $M$ denotes the number
of independent check nodes contained in~$\mathcal{G}$.

We now prove the case that $\mathcal{G}$ is a disjoint graph. Let
$\mathcal{G}$ contain $p$~connected subgraphs: $\mathcal{G}_1,
\mathcal{G}_2, \ldots, \mathcal{G}_p$. By the above analysis, the
complexity of encoding~$\mathcal{G}_i$ is $\mathcal{O}(M_i),
i=1,2,\ldots,p$, where $M_i$ denotes the number of independent
check nodes contained in~$\mathcal{G}_i$. Since $\mathcal{G} =
\mathcal{G}_1 \cup \mathcal{G}_2 \cup \ldots \cup \mathcal{G}_p$,
then the encoding complexity of $\mathcal{G}$ is $\sum_{i=1}^p
\mathcal{O}(M_i) = \mathcal{O}(\sum_{i=1}^p M_i) = \mathcal{O}(M)$
where $M$ is the number of independent check nodes in
$\mathcal{G}$. This completes the proof. \hfill $\Box$

From Theorem~\ref{thm:theorem2}, we easily derive the following
corollaries.
\begin{corollary}
\label{thm:corollary1} If a Tanner graph does not contain any
encoding stopping set, then it can be represented by a union of
pseudo-trees.
\end{corollary}
The proof of Corollary~\ref{thm:corollary1} can be found in the
proof of Theorem~\ref{thm:theorem2}.

\begin{corollary}
\label{thm:corollary4} The label-and-decide algorithm can encode
any tree LDPC codes (whose Tanner graphs are cycle-free) with
linear complexity.
\end{corollary}
\emph{Proof}: Let $\mathcal{T}$ be the Tanner graph of a tree LDPC
code and $\mathcal{S}$ be an arbitrary subgraph of~$\mathcal{T}$.
Since the Tanner graph~$\mathcal{T}$ is a tree, its
subgraph~$\mathcal{S}$ is either a tree or a union of trees.
Therefore, the graph~$\mathcal{S}$ contains at least one bit leaf
node with degree one. Since the graph~$\mathcal{S}$ contains a
degree-one bit node, $\mathcal{S}$ can not be an encoding stopping
set. Since no subgraph of~$T$ is an encoding stopping set, by
Theorem~\ref{thm:theorem2}, the tree code~$\mathcal{T}$ can be
encoded in linear complexity by the label-and-decide algorithm.
This completes the proof. \hfill $\Box$

\begin{corollary}
\label{thm:corollary2} A regular LDPC code with column weight~2
(cycle code) can be encoded in linear complexity by the
label-and-decide algorithm.
\end{corollary}
\emph{Proof}: We prove Corollary~\ref{thm:corollary2} by showing
that a cycle code does not contain any encoding stopping set.
Assume the cycle code contains an encoding stopping
set~$\mathcal{E_s}$. By the definition of cycle code and
condition~(B2), all the bit nodes in~$\mathcal{E_s}$ have uniform
degree two. It follows that the binary sum of all the parity check
equations in~$\mathcal{E_s}$ is a vector of 0's. Then, at least
one check node in~$\mathcal{E_s}$ is dependent on the other check
nodes. This contradicts condition~(B3) that all the check nodes in
an encoding stopping set are independent of each other. Hence, a
cycle code does not contain any encoding stopping set. By
Theorem~\ref{thm:theorem2}, a cycle code is linear time encodable
by the label-and-decide algorithm. This completes the proof.
\hfill $\Box$

An alternative proof can be found in~\cite{Jin:1}.
\begin{figure}[htb]
\centerline{\epsfig{figure=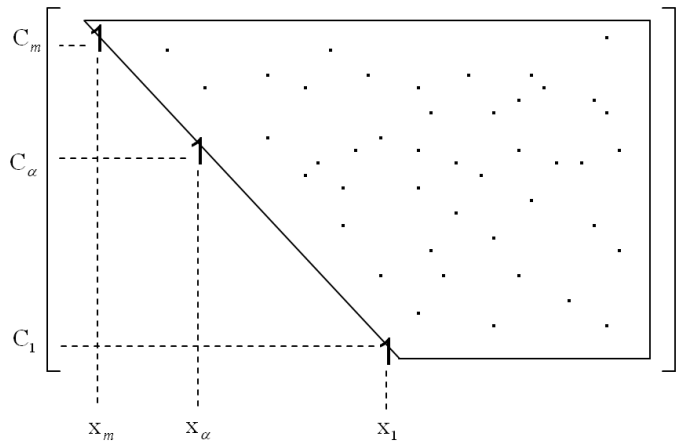,width=9cm,height=6cm}}
\caption{\label{encode6} A parity check matrix in upper triangular
form.}
\end{figure}
\begin{corollary}
\label{thm:corollary3} Let $\bf H$ be the parity check matrix of
an LDPC code. If $\bf H$ can be transformed into an upper
triangular matrix $\bf U$ by row and column permutations, then the
LDPC code can be encoded in linear time by the label-and-decide
algorithm.
\end{corollary}
\emph{Proof}: We label  the rows of the upper triangular
matrix~$\bf U$ one by one as $C_1$, $C_2$, $\ldots$, $C_m$, from
the bottom to the top, as shown in Figure~\ref{encode6}. We notice
that if $i > j$, then there exists at least one bit that is
contained in~$C_i$ but not in~$C_j$. Assume the Tanner graph of
the code contains an encoding stopping set~$\mathcal{E_s}$ that
contains check nodes $C_{i_1}$, $C_{i_2}$, $\ldots$, $C_{i_p}$.
Let $q = \max(i_1, i_2, \ldots, i_p)$. There exists at least one
bit node~$x_q$ in~$\mathcal{E_s}$ that only connects to~$C_q$.
This contradicts the fact that every bit node in an encoding
stopping set is connected to at least two check nodes in the
encoding stopping set. Hence, $\mathcal{E_s}$ is not an encoding
stopping set. Since the Tanner graph of the LDPC code does not
contain any encoding stopping set, by Theorem~\ref{thm:theorem2}
it is linear time encodable. This completes the proof. \hfill
$\Box$

Theorems~\ref{thm:theorem1} and~\ref{thm:theorem2} show that
encoding stopping sets prevent the application of the
label-and-decide algorithm. However, we will show in the next
section that encoding stopping sets can also be encoded in linear
complexity.

\section{Linear complexity encoding approach for encoding stopping sets}
\label{sec:linear-encoding-stoppingset}
Let $\mathcal{E_s}$ be an
encoding stopping set. We say $\mathcal{E_s}$ is a
\emph{k-fold-constraint encoding stopping set} if the following
two conditions hold.
\begin{itemize}
\item[(C1)] There exist $k$~check nodes $C_1$, $C_2$, $\ldots$,
$C_k$ in~$\mathcal{E_s}$ such that $\mathcal{E_s} \backslash
\{C_1, C_2, \ldots, C_k \}$ does not contain any encoding stopping
set. We call the $k$~check nodes $C_1$, $C_2$, $\ldots$, $C_k$
\emph{key check nodes}. \item[(C2)] For any $k-1$~check nodes
$C_1$, $C_2$, $\ldots$, $C_{k-1}$ in~$\mathcal{E_s}$,
$\mathcal{E_s} \backslash \{C_1, C_2, \ldots, C_{k-1} \}$ contains
an encoding stopping set.
\end{itemize}
The notation $\mathcal{E_s} \backslash \{C_1, C_2, \ldots, C_k \}$
denotes the remaining graph after deleting check nodes $C_1$,
$C_2$, $\ldots$, $C_k$ from $\mathcal{E_s}$. Figure~\ref{encode4}
shows a 2-fold-constraint encoding stopping set. After deleting
the two key check nodes~$C_8$ and~$C_9$ from this encoding
stopping set, the Tanner graph turns into a pseudo-tree, see
Figure~\ref{encode2}.  We will focus on 1-fold-constraint and
2-fold-constraint encoding stopping sets in this paper, since we
will show later that all types of LDPC codes can be decomposed
into 1-fold or 2-fold constraint encoding stopping sets and
pseudo-trees.

Let us first look at a 2-fold-constraint encoding stopping
set~$\mathcal{E_s}$ of size~$M$. By definition, there exist two
key check nodes~$C_\alpha$ and~$C_\beta$ in~$\mathcal{E_s}$ such
that $\mathcal{E_s} \backslash \{C_\alpha, C_\beta\}$ does not
contain any encoding stopping set. We encode $\mathcal{E_s}$ in
three steps. In the first step, we encode $\mathcal{E_s}
\backslash \{C_\alpha, C_\beta\}$ using the label-and-decide
algorithm according to Theorem~\ref{thm:theorem2}. During
encoding, $M-2$~bit nodes are labeled as parity bits and the
remaining bit nodes are labeled as information bits. In the second
step, we verify the two key check nodes~$C_\alpha$ and~$C_\beta$
based on the bit values acquired in Step~1. The key check
nodes~$C_\alpha$ and~$C_\beta$ indicate that two bits~$x_\gamma$
and~$x_\delta$ that were previously labeled as information bits
are actually parity bits, and their values are determined
by~$C_\alpha$ and~$C_\beta$. We call the two bits~$x_\gamma$
and~$x_\delta$ \emph{reevaluated bits}. The bits $x_\gamma$ and
$x_\delta$ satisfy the following three conditions.
\begin{itemize}
\item[(D1)] $x_\gamma$ is constrained by the parity check
equation~$C_\alpha$. \item[(D2)] $x_\delta$ is constrained by the
parity check equation~$C_\beta$. \item[(D3)] $x_\gamma$ and
$x_\delta$ are not both contained in $C_\alpha$ and $C_\beta$.
\end{itemize}
Since $C_\alpha$, $C_\beta$, and the other check nodes in
$\mathcal{E_s}$ are independent of each other, there must exist
bit nodes~$x_\gamma$ and~$x_\delta$ that satisfy conditions~(D1)
to~(D3). An algorithm for finding reevaluated bits~$x_\gamma$
and~$x_\delta$ from a 2-fold-constraint encoding stopping set is
presented in Appendix A. Notice that the work load to find the two
key check nodes and the two reevaluated bits is a preprocessing
step that is carried out only once. Assume in step~1 that
$x_\gamma$ and~$x_\delta$ are randomly assigned initial
values~$x_\gamma^0$ and $x_\delta^0$, respectively. If the parity
check equations $C_\alpha$ and $C_\beta$ are both satisfied, the
initial values~$x_\gamma^0$ and $x_\delta^0$ are the correct
values for~$x_\gamma$ and~$x_\delta$. If $C_\alpha$, or $C_\beta$,
or both, are not satisfied, we need to recompute the values of
$x_\gamma$ and $x_\delta$ from the values of the key check
nodes~$C_\alpha$ and~$C_\beta$. Let $\triangle x_\gamma =
\widetilde{x_\gamma} - x_\gamma^0$ and $\triangle x_\delta =
\widetilde{x_\delta} - x_\delta^0$ where $\widetilde{x_\gamma}$
and $\widetilde{x_\delta}$ are the correct values of~$x_\gamma$
and~$x_\delta$, respectively, and let~$\widetilde{C_\alpha}$
and~$\widetilde{C_\beta}$ be the values of the key check nodes
$C_\alpha$ and $C_\beta$, respectively. If $x_\gamma$ is contained
in both $C_\alpha$ and $C_\beta$, and $x_\delta$ is only contained
in $C_\beta$, we derive the following equations.
\begin{equation}
\label{equ1}
\begin{array}{c}
\begin{array}{ccccc}
                  \triangle x_\gamma &=& -\widetilde{C_\alpha} = \widetilde{C_\alpha} \\
                  \triangle x_\gamma \oplus \triangle x_\delta &=&
                  -\widetilde{C_\beta} = \widetilde{C_\beta}
                             \end{array}    \end{array}
\end{equation}
If $x_\delta$ is contained in both $C_\alpha$ and $C_\beta$, and
$x_\gamma$ is only contained in $C_\alpha$, we derive the
following equations.
\begin{equation}
\label{equ2}
\begin{array}{c}
\begin{array}{ccccc}
                  \triangle x_\gamma \oplus \triangle x_\delta &=&
                  -\widetilde{C_\alpha} = \widetilde{C_\alpha} \\
                  \triangle x_\delta &=& -\widetilde{C_\beta} = \widetilde{C_\beta}

                             \end{array}    \end{array}
\end{equation}
If $x_\gamma$ is only contained in $C_\alpha$ and $x_\delta$ is
only contained in $C_\beta$, we have the following equations.
\begin{equation}
\label{equ3}
\begin{array}{c}
\begin{array}{ccccc}
                  \triangle x_\gamma &=&
                  -\widetilde{C_\alpha} = \widetilde{C_\alpha} \\
                  \triangle x_\delta &=& -\widetilde{C_\beta} = \widetilde{C_\beta}

                             \end{array}    \end{array}
\end{equation}
From equations~(\ref{equ1}) to~(\ref{equ3}), we can get the
correct values of~$x_\gamma$ and~$x_\delta$. In the third step, we
recompute those parity bits that are affected by the new values
of~$x_\gamma$ and~$x_\delta$. This encoding method is named
\emph{label-decide-recompute} and is described in
Algorithm~\ref{encoding-algorithm2}.
\begin{algorithm}
\caption{Label-decide-recompute algorithm for a 2-fold-constraint
encoding stopping set~$\mathcal{E_s}$ of size~$M$
\label{encoding-algorithm2}}
\begin{algorithmic}
\STATE {\bf Preprocessing (carry out only once):} \STATE Find two
check nodes $C_\alpha$ and $C_\beta$ such that $\mathcal{E_s}
\backslash \{ C_\alpha, C_\beta \}$ does not contain any encoding
stopping set; \STATE Using Algorithm~\ref{encoding-algorithm8} to
pick two information bits~$x_\gamma$ and $x_\delta$ that satisfy
conditions~(D1) to~(D3).
 \STATE Determine the parity bits $x_{p_1}$,
$x_{p_2}$, $\ldots$, $x_{p_s}$ that are affected by the values
of~$x_\gamma$ and~$x_\delta$; \STATE {\bf Encoding:} \STATE Fill
the values of the information bits except for~$x_\gamma$
and~$x_\delta$; \STATE Assign $x_\gamma = 0$ and $x_\delta = 0$;
\STATE Encode $\mathcal{E_s} \backslash \{ C_\alpha, C_\beta \}$
using Algorithm~\ref{encoding-algorithm1}. Compute the values of
the $M-2$~parity bits; \STATE Compute the values
$\widetilde{C_\alpha}$ and $\widetilde{C_\beta}$ of the key check
nodes $C_\alpha$ and $C_\beta$, respectively; \IF
{$\widetilde{C_\alpha} \neq 0$ or $\widetilde{C_\beta} \neq 0$}
    \STATE Recompute the values of~$x_\gamma$ and~$x_\delta$ from~$\widetilde{C_\alpha}$ and~$\widetilde{C_\beta}$ by
    equations~(\ref{equ1}) to~(\ref{equ3});
    \FOR{$i$ = 1 to $s$}
        \STATE Recompute the value of the parity bit~$x_{p_i}$ based on
            the new values of $x_\gamma$ and $x_\delta$;
    \ENDFOR
\ENDIF \STATE Output the encoding result.
\end{algorithmic}
\end{algorithm}

Next, we analyze the computation complexity of the
label-decide-recompute algorithm when encoding a 2-fold-constraint
encoding stopping set. Every check node except for the two key
check nodes $C_\alpha$ and $C_\beta$ are computed at most twice in
the label-decide-recompute encoding (label-and-decide step and
recompute step) while the two key check nodes $C_\alpha$ and
$C_\beta$ need to be computed only once. In addition, we need one
extra XOR operation to compute the two reevaluated bits $x_\gamma$
and $x_\delta$ by equations~(\ref{equ1}) to~(\ref{equ3}). Hence,
the encoding complexity of the label-decide-recompute algorithm is
less than or equal to $2 \cdot \sum_{i=1}^{M-2} (k_i - 2) +
(k_\alpha - 1) + (k_\beta - 1) + 1$ where $k_i$, $1 \leq i \leq
M-2$, are the degrees of the check nodes other than $C_\alpha$,
$C_\beta$ and $k_\alpha$, $k_\beta$ are the degrees of the check
nodes $C_\alpha$ and $C_\beta$, respectively. The encoding
complexity of the label-decide-recompute algorithm can be further
simplified to be less than $2 \cdot M \cdot (\overline{k}-1)$
where $M$ is the number of check nodes in the encoding stopping
set and $\overline{k}$ is the average number of bit nodes
contained in each check node in the encoding stopping set. This
shows that the label-decide-recompute algorithm encodes any
2-fold-constraint encoding stopping set in linear time. The
pre-processing (determining key check nodes, reevaluated bits, and
parity bits affected by the reevaluated bits) is done offline and
does not count towards encoder complexity.

We look at an example. Figure~\ref{encode4} shows a
2-fold-constraint encoding stopping set~$\mathcal{E_s}$. After
deleting the two check nodes~$C_8$ and~$C_9$, $\mathcal{E_s}$
becomes the pseudo-tree shown in Figure~\ref{encode2}. In
addition, the value of the bit node~$x_5$ affects~$C_8$ and the
value of the bit node~$x_8$ affects~$C_9$. Hence, the two bits
$x_5$ and $x_8$ are reevaluated bits. We use the
label-decide-recompute algorithm to encode~$\mathcal{E_s}$ as
follows.
\begin{itemize}
\item[Step 1.] Assign $x_5 = 0$ and $x_8 = 0$. Encode the
pseudo-tree part following the procedures on page~10. \item[Step
2.] Compute the values of the key check nodes~$C_8$ and~$C_9$,
e.g., $\widetilde{C_8} = x_1 \oplus x_2 \oplus x_3 \oplus x_4$ and
$\widetilde{C_9} = x_1 \oplus x_2 \oplus x_{8} \oplus x_{16}$.
\begin{itemize}
\item[Step 3a.]  If $\widetilde{C_8} = 0$ and~$\widetilde{C_9} =
0$, stop encoding and output the codeword $[x_1 \,\, x_2 \,\,
\ldots \,\, x_{16}]$. \item[Step 3b.] If $\widetilde{C_8} = 1$
or~$\widetilde{C_9} = 1$, recompute the values of~$x_5$ and $x_8$
as follows: $x_5 = \widetilde{C_8}$ and $x_8 = \widetilde{C_9}$,
where $\widetilde{C_8}$ and $\widetilde{C_9}$ are the values of
the parity check equations~$C_8$ and~$C_9$, respectively.
Recompute the parity bits~$x_1$, $x_2$, $x_3$, and $x_4$ based on
the new values of~$x_5$ and~$x_8$. Output the codeword $[x_1 \,\,
x_2 \,\, \ldots \,\, x_{16}]$.
\end{itemize}
\end{itemize}

The label-decide-recompute algorithm can be further simplified. We
restudy the third step of the label-decide-recompute method.
Assume $p_1$, $p_2$, $\ldots$, $p_m$ are the parity bits whose
values need to be updated. In order to get the new values of the
parity bits $p_1$, $p_2$, $\ldots$, $p_m$, we need to recompute
those parity check equations that relate to $p_1$, $p_2$,
$\ldots$, $p_m$. In fact, instead of recomputing the parity check
equations relating to $p_1$, $p_2$, $\ldots$, $p_m$, we can
directly flip the values of the parity bits $p_1$, $p_2$,
$\ldots$, $p_m$ since in the binary field the value of a bit is
either~0 or~1. For example, if the correct value of~$x_\gamma$ is
different from its original value~$x_\gamma^0$, we simply flip the
values of those parity bits that are affected by~$x_\gamma$. We
name the above encoding method \emph{label-decide-flip} and
describe it in Algorithm~\ref{encoding-algorithm3}. The encoding
complexity of Algorithm~\ref{encoding-algorithm3} is $M \cdot
(\overline{k}-1)$ XOR operations plus two vector flipping
operations.
\begin{algorithm}
\caption{Label-decide-flip algorithm for a 2-fold-constraint
encoding stopping set~$\mathcal{E_s}$ of size~$M$
\label{encoding-algorithm3}}
\begin{algorithmic}
\STATE {\bf Preprocessing (carry out only once):} \STATE Find two
check nodes $C_\alpha$ and $C_\beta$ such that $\mathcal{E_s}
\backslash \{ C_\alpha, C_\beta \}$ does not contain any encoding
stopping set; \STATE Using Algorithm~\ref{encoding-algorithm8} to
pick two information bits~$x_\gamma$ and $x_\delta$ that satisfy
conditions (D1) to (D3). \STATE Determine the parity bits
$x_{u_1}$, $x_{u_2}$, $\ldots$, $x_{u_r}$ that are affected by the
value of~$x_\gamma$ and group $x_{u_1}$, $x_{u_2}$, $\ldots$,
$x_{u_r}$ in a vector $\overrightarrow{X_\gamma} = [x_{u_1} \,\,
x_{u_2} \,\, \ldots \,\, x_{u_r}]$. Determine the parity bits
$x_{p_1}$, $x_{p_2}$, $\ldots$, $x_{p_s}$ that are affected by the
value of~$x_\delta$ and group $x_{p_1}$, $x_{p_2}$, $\ldots$,
$x_{p_s}$ in a vector $\overrightarrow{X_\delta} = [x_{p_1} \,\,
x_{p_2} \,\, \ldots \,\, x_{p_s}]$. Determine the parity bits
$x_{q_1}$, $x_{q_2}$, $\ldots$, $x_{q_t}$ that are affected by the
value of~$x_\gamma \oplus x_\delta$ and group $x_{q_1}$,
$x_{q_2}$, $\ldots$, $x_{q_t}$ in a vector
$\overrightarrow{X_\epsilon} = [x_{q_1} \,\, x_{q_2} \,\, \ldots
 \,\, x_{q_t}]$.  \STATE {\bf Encoding:}
\STATE Fill the values of the information bits except
for~$x_\gamma$ and~$x_\delta$; \STATE Assign $x_\gamma = 0$ and
$x_\delta = 0$; \STATE Encode $\mathcal{E_s} \backslash \{
C_\alpha, C_\beta \}$ using Algorithm~\ref{encoding-algorithm1}.
Compute the values of the $M-2$~parity bits; \STATE Compute the
values $\widetilde{C_\alpha}$ and $\widetilde{C_\beta}$ of the
parity check equations $C_\alpha$ and $C_\beta$, respectively; \IF
{$\widetilde{C_\alpha} \neq 0$ or $\widetilde{C_\beta} \neq 0$}
    \STATE Recompute the values of~$x_\gamma$ and~$x_\delta$ from~$\widetilde{C_\alpha}$ and~$\widetilde{C_\beta}$ by
    equations~(\ref{equ1}) to~(\ref{equ3});
    \IF{$x_\gamma = 1$}
    \STATE Flip the vector $\overrightarrow{X_\gamma}$.
    \ENDIF
    \IF{$x_\delta = 1$}
    \STATE Flip the vector $\overrightarrow{X_\delta}$.
    \ENDIF
    \IF{$x_\gamma \oplus x_\delta = 1$}
    \STATE Flip the vector $\overrightarrow{X_\epsilon}$.
    \ENDIF
\ENDIF \STATE Output the encoding result.
\end{algorithmic}
\end{algorithm}
We, again, look at an example. The 2-fold-constraint encoding
stopping set shown in Figure~\ref{encode4} can be encoded by
Algorithm~\ref{encoding-algorithm3} as follows.
\begin{itemize}
\item[Step 1.] Assign $x_5 = 0$ and $x_8 = 0$. Encode the
pseudo-tree part following the procedures on page~10. \item[Step
2.] Compute the values of the parity check equations~$C_8$
and~$C_9$, e.g., $\widetilde{C_8} = x_1 \oplus x_2 \oplus x_3
\oplus x_4$ and $\widetilde{C_9} = x_1 \oplus x_2 \oplus x_{8}
\oplus x_{16}$.
\begin{itemize}
\item[Step 3a.]  If $\widetilde{C_8} = 0$ and~$\widetilde{C_9} =
0$, stop encoding and output the codeword $[x_1 \,\, x_2 \,\,
\ldots \,\, x_{16}]$. \item[Step 3b.] If $\widetilde{C_8} = 1$
or~$\widetilde{C_9} = 1$, recompute the values of~$x_5$ and $x_8$
as the following: $x_5 = \widetilde{C_8}$ and $x_8 =
\widetilde{C_9}$ where $\widetilde{C_8}$ and $\widetilde{C_9}$ are
the values of the parity check equations~$C_8$ and~$C_9$,
respectively. If $x_5 = 1$, flip the vector $[x_1 \,\, x_2 \,\,
x_3]$ to be $[\sim x_1 \,\, \sim x_2 \,\, \sim x_3]$. If $x_8 =
1$, flip the vector $[x_3 \,\, x_4]$. If $x_5 \oplus x_8 = 1$,
flip the vector $[x_4]$. Output the codeword $[x_1 \,\, x_2 \,\,
\ldots \,\, x_{16}]$.
\end{itemize}
\end{itemize}

It is easy to revise Algorithm~\ref{encoding-algorithm2} and
Algorithm~\ref{encoding-algorithm3} to encode a 1-fold-constraint
encoding stopping set. For example,
Algorithm~\ref{encoding-algorithm4} shows the
label-decide-recompute algorithm for a 1-fold-constraint encoding
stopping set. The encoding complexity of
Algorithm~\ref{encoding-algorithm4} is less than~$2 \cdot M \cdot
(\overline{k}-1)$ where $M$ is the number of check nodes in the
encoding stopping set and $\overline{k}$ is the average number of
bit nodes contained in each check node in the encoding stopping
set.

\begin{algorithm}
\caption{Label-decide-recompute algorithm for a 1-fold-constraint
encoding stopping set $\mathcal{E_s}$ of size~$M$
\label{encoding-algorithm4}}
\begin{algorithmic}
\STATE {\bf Preprocessing (carry out only once):} \STATE Find a
check node~$C^*$ such that $\mathcal{E_s} \backslash C^*$ does not
contain any encoding stopping set. \STATE Pick an information
bit~$x^*$ that is constrained by the parity check equation~$C^*$.
\STATE Determine the parity bits $x_{p_1}$, $x_{p_2}$, $\ldots$,
$x_{p_s}$ that are affected by~$x^*$. \STATE {\bf Encoding:}
\STATE Fill the values of the information bits except for~$x^*$.
\STATE Assign $x^* = 0$. \STATE Encode $\mathcal{E_s} \backslash
C^*$ using Algorithm~\ref{encoding-algorithm1}, compute the values
of the $M-1$~parity bits. \STATE Verify the parity check
equation~$C^*$. \IF {the parity check equation $C^*$ is not
satisfied}
    \STATE $x^* \leftarrow 1$.
    \FOR{$i$ = 1 to s}
        \STATE Recompute the value of the parity bit $x_{p_i}$ based on
            the new value of $x^*$;
    \ENDFOR
\ENDIF \STATE Output the encoding result.
\end{algorithmic}
\end{algorithm}

\section{Linear complexity encoding for general LDPC codes}
\label{sec:linear-encoding-ldpccodes}
In this section, we propose
a linear complexity encoding method for general LDPC codes. We
will show that any Tanner graph can be decomposed into
pseudo-trees and encoding stopping sets that are 1-fold-constraint
or 2-fold-constraint. By encoding each pseudo-tree or encoding
stopping set using Algorithm~\ref{encoding-algorithm1} or
Algorithm~\ref{encoding-algorithm2}, we achieve linear time
encoding for arbitrary LDPC codes.

\begin{figure}[htb]
\centerline{\epsfig{figure=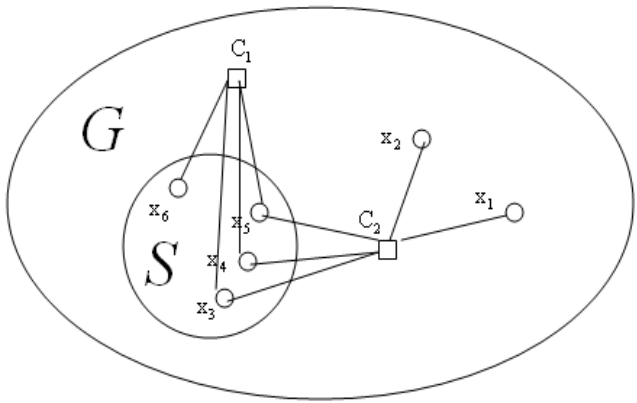,width=9cm,height=6cm}}
\caption{\label{encode29} Outsider nodes.}
\end{figure}

To proceed, we provide the following definition. Given a Tanner
graph~$\mathcal{G}$ and its subgraph~$\mathcal{S}$, we call the
bit nodes in~$\mathcal{G}$ but not~in $\mathcal{S}$ the
\emph{outsider nodes of~$\mathcal{S}$}. For example,
Figure~\ref{encode29} shows a Tanner graph~$\mathcal{G}$ and its
subgraph~$\mathcal{S}$. Since in Figure~\ref{encode29} the two bit
nodes~$x_1$ and~$x_2$ are in~$\mathcal{G}$ but not
in~$\mathcal{S}$, $x_1$ and $x_2$ are outsider nodes
of~$\mathcal{S}$. The check node $C_2$ contains two outsider nodes
of~$\mathcal{S}$, i.e., is connected to two outsider nodes. The
check node $C_1$ contains zero outsider nodes of~$\mathcal{S}$.

We start from LDPC codes with maximum column weight~3 by proving
the following lemma.
\begin{lemma}
\label{lemma2} Assume the maximum bit node degree of a Tanner
graph~$\mathcal{G}$ is three, then one of the following statements
must be true.
\begin{itemize}
\item[(E1)] There are no pseudo encoding stopping sets or encoding
stopping sets in~$\mathcal{G}$.
\item[(E2)] There exists a pseudo
encoding stopping set in $\mathcal{G}$. All the bit nodes in the
pseudo encoding stopping set have uniform degree~2.
\item[(E3)]
There exists a 1-fold-constraint or a 2-fold-constraint encoding
stopping set in $\mathcal{G}$.
\end{itemize}
\end{lemma}
\emph{Proof}: We only need to prove either condition (E2), or
condition (E3), is true if $\mathcal{G}$ contains a pseudo
encoding stopping set, or an encoding stopping set, respectively.
We prove this statement by constructing a subgraph~$\mathcal{S}$
from the Tanner graph~$\mathcal{G}$. Initially $\mathcal{S}$ is
empty. We pick a check node~$C_1$ from~$\mathcal{G}$ that contains
the smallest number of bit nodes. Next, we add $C_1$ and all the
bit nodes contained in $C_1$ to~$\mathcal{S}$. We keep adding
check nodes and their associated bit nodes to $\mathcal{S}$ till
$\mathcal{S}$ contains a pseudo encoding stopping set or an
encoding stopping set. Each time we add a check node
to~$\mathcal{S}$, we always pick the check node that contains the
fewest outsider nodes of~$\mathcal{S}$. If $\mathcal{S}$ contains
an encoding stopping set, we also add all the check nodes in
$\mathcal{G} \backslash \mathcal{S}$ that contain zero outsider
nodes to $\mathcal{S}$. Next, we discuss two different cases.

{\bf $\mathcal{S}$ contains an encoding stopping set.} Assume
$\mathcal{S}$ contains $k$ check nodes and the $j$-th added check
node $C_j$ is the last check node that introduces outsider nodes
to $\mathcal{S}$. We will show that $k - j \leq 2$. Assume $C_j$
adds $p$ outsider nodes $x_{j_1}$, $x_{j_2}$, $\ldots$, $x_{j_p}$
to $\mathcal{S}$. We will prove that the $(j+1)$-th added check
node $C_{j+1}$ connects to all the $p$ bit nodes $x_{j_1}$,
$x_{j_2}$, $\ldots$, $x_{j_p}$. If $C_{j+1}$ does not connect to
all the $p$ bit nodes, then $C_{j+1}$ contains a smaller number of
outsider nodes than~$C_j$ and should be added earlier than $C_j$
since we always pick the check node that contains the smallest
number of outsider nodes and add it first to $\mathcal{S}$. This
contradicts the fact that $C_{j+1}$ is added to $\mathcal{S}$
after~$C_j$. Therefore, $C_{j+1}$ should connect to all the $p$
bit nodes $x_{j_1}$, $x_{j_2}$, $\ldots$, $x_{j_p}$. Similarly,
$C_{j+2}$, $\ldots$, $C_k$ connect to all the $p$ bit nodes
$x_{j_1}$, $x_{j_2}$, $\ldots$, $x_{j_p}$. Since any bit node can
connect to at most three check nodes, it follows that $k - j \leq
2$, which means at most two check nodes are added to~$\mathcal{S}$
after $C_j$. Notice that $\mathcal{S}$ does not contain any
encoding stopping set before adding~$C_{j+1}$. Hence, the encoding
stopping set in~$\mathcal{S}$ is either a 1-fold-constraint
encoding stopping set or a 2-fold-constraint encoding stopping
set. Condition~(E2) is satisfied.

{\bf $\mathcal{S}$ is a pseudo encoding stopping set.} It follows
that the binary sum of all the check nodes in~$\mathcal{S}$ is
zero. So, the degree of every bit node in~$\mathcal{S}$ is an even
number. Since the maximum bit node degree is three, the degree of
each bit node in~$\mathcal{S}$ is two. Condition (E3) is
satisfied.

This completes the proof. \hfill $\Box$

We detail the method of determining a pseudo encoding stopping set
or an encoding stopping set in
Algorithm~\ref{encoding-algorithm9}.

\begin{algorithm}
\caption{Find a pseudo encoding stopping set or an encoding
stopping set (1-fold-constraint or 2-fold-constraint) from a
Tanner graph $\mathcal{G}$ with maximum bit node degree~3.
\label{encoding-algorithm9}}
\begin{algorithmic}
\STATE $\mathcal{S} = \phi$. \STATE $Flag \leftarrow 0$. \STATE $i
= 1$. \WHILE {$Flag = 0$ and $\mathcal{G} \neq \mathcal{S}$}
       \STATE Find a check node $C_i$ in $\mathcal{G} \backslash \mathcal{S}$ that
       contains the smallest number of outsider nodes of $\mathcal{S}$.
       \STATE Add $C_i$ and all its associated outsider nodes to
       $\mathcal{S}$.
       \IF {$C_i$ does not introduce new bit nodes to $\mathcal{S}$}
           \STATE $\mathcal{A} \leftarrow \mathcal{S}$.
           \WHILE{there exists a bit node $x$ of degree one in $\mathcal{A}$}
      \STATE Delete the degree-one bit node $x$ and the check
      node connecting to $x$ from $\mathcal{A}$.
\ENDWHILE \IF {$\mathcal{A} = \phi$}
    \STATE $\mathcal{S}$ does not contain any pseudo encoding stopping set or
encoding stopping set. \ELSE
    \STATE $Flag \leftarrow 1$.
\ENDIF
       \ENDIF
       \STATE $i = i + 1$.
\ENDWHILE \IF {$Flag = 1$}
    \IF {all the bit nodes in $\mathcal{A}$ are of degree 2}
    \STATE The subgraph $\mathcal{A}$ is a pseudo encoding
    stopping set.
    \ELSE
    \STATE The subgraph $\mathcal{A}$ is an encoding stopping set.
    \ENDIF
    \STATE Output $\mathcal{A}$.
    \ELSE
    \STATE The Tanner graph $\mathcal{G}$ does not contain
    pseudo encoding stopping sets or encoding stopping sets.
    \ENDIF
\end{algorithmic}
\end{algorithm}

\begin{theorem}
\label{thm:theorem3} Let $\mathcal{G}$ be the Tanner graph of an
LDPC code. If the maximum bit node degree of~$\mathcal{G}$ is
three, then the LDPC code can be encoded in linear time and the
encoding complexity is less than $2 \cdot M \cdot (\overline{k} -
1)$ where $M$ is the number of independent check nodes in
$\mathcal{G}$ and $\overline{k}$ is the average number of bit
nodes contained in each check node.
\end{theorem}
\emph{Proof}: If the Tanner graph $\mathcal{G}$ does not contain
any encoding stopping set, then the corresponding LDPC code can be
encoded in linear time by Theorem~\ref{thm:theorem2}. Therefore,
we only need to prove Theorem~\ref{thm:theorem3} for the case that
$\mathcal{G}$ contains encoding stopping sets. Since the maximum
bit node degree of~$\mathcal{G}$ is three, by Lemma~\ref{lemma2}
there exists a pseudo encoding stopping set or an encoding
stopping set~$\mathcal{G}_1$ in $\mathcal{G}$. If $\mathcal{G}_1$
is a pseudo encoding stopping set, we simply delete a redundant
check node from~$\mathcal{G}_1$ and $\mathcal{G}_1$ becomes a
pseudo-tree. If $\mathcal{G}_1$ is an encoding stopping set, it is
either a 1-fold-constraint or a 2-fold-constraint encoding
stopping set by Lemma~\ref{lemma2}.

Next, we look at the subgraph~$\mathcal{G} \backslash
\mathcal{G}_1$. We first transform the parity check equations
in~$\mathcal{G} \backslash \mathcal{G}_1$ into generalized parity
check equations by moving the bits contained in~$\mathcal{G}_1$
from the left-hand side of the equation to the right-hand side of
the equation. Let a parity check equation $C$ contain $k$~bit
nodes $x_1$, $x_2$, $\ldots$, $x_k$ where the bits $x_{q+1}$,
$\ldots$, $x_k$ are also in~$\mathcal{G}_1$, then the parity check
equation~$C$ can be rewritten as
\begin{equation}
\label{equ5} x_1 \oplus x_2 \oplus \ldots \oplus x_k = 0
\Longrightarrow x_1 \oplus x_2 \oplus \ldots \oplus x_q = x_{q+1}
\oplus x_{q+2} \oplus \ldots \oplus x_k = b
\end{equation}
In equation~(\ref{equ5}), $b$ becomes a constant after we
encode~$\mathcal{G}_1$ and get the values of all the bits
in~$\mathcal{G}_1$. Since the maximum bit node degree
of~$\mathcal{G} \backslash \mathcal{G}_1$ is less than or equal to
three, we, again, find a pseudo encoding stopping set or an
encoding stopping set~$\mathcal{G}_2$ from~$\mathcal{G} \backslash
\mathcal{G}_1$. If $\mathcal{G}_2$ is an encoding stopping set,
$\mathcal{G}_2$ is either a 1-fold-constraint or a
2-fold-constraint encoding stopping set by Lemma~\ref{lemma2}. If
$\mathcal{G}_2$ is a pseudo encoding stopping set and we assume
$\mathcal{G}_2$ contains the following $m$~generalized parity
check equations,
\begin{equation}
\label{equ6} \begin{array}{c}
\begin{array}{ccccccccc}
                  x_{1,1} & \oplus & x_{1,2} & \oplus & \ldots & x_{1,a_1} & = & b_1 \\
                  x_{2,1} & \oplus & x_{2,2} & \oplus & \ldots & x_{2,a_2} & = & b_2 \\
                  \vdots &  & \vdots &  &  & \vdots &  & \vdots \\
                  x_{m,1} & \oplus & x_{m,2} & \oplus & \ldots & x_{m,a_m} & = &
                  b_m
                  \end{array}    \end{array}
\end{equation}
we derive that
\begin{equation}
\label{equ8} b_1 \oplus b_2 \oplus \ldots \oplus b_m = 0
\end{equation}
Hence, we can replace any generalized parity check equation
in~(\ref{equ6}) by the new parity check equation~(\ref{equ8}).
From the above analysis, we can delete any check node from
$\mathcal{G}_2$ to make $\mathcal{G}_2$ a pseudo-tree. To maintain
the code structure, we also generate a new check node $C^*$ that
represents the parity check equation~(\ref{equ8}). Since the
parity check equation~(\ref{equ8}) only contains bits
in~$\mathcal{G}_1$, we add the new check node~$C^*$
to~$\mathcal{G}_1$ and regenerate encoding stopping sets or
pseudo-trees in the graph~$\mathcal{G}_1 \cup C^*$.

Generally, we can find a pseudo encoding stopping set or an
encoding stopping set~$\mathcal{G}_{i+1}$ from the
subgraph~$\mathcal{G} \backslash \{\mathcal{G}_1 \cup
\mathcal{G}_2 \cup \ldots \cup \mathcal{G}_i\}$. If
$\mathcal{G}_{i+1}$ is an encoding stopping set,
$\mathcal{G}_{i+1}$ is either a 1-fold-constraint or a
2-fold-constraint encoding stopping set by Lemma~\ref{lemma2}. If
$\mathcal{G}_{i+1}$ is a pseudo encoding stopping set, we operate
in three steps. In the first step, we sum up all the generalized
parity check equations in $\mathcal{G}_{i+1}$ to generate a new
parity check equation $C^*$. In the second step, we delete one
check node from $\mathcal{G}_{i+1}$ to make $\mathcal{G}_{i+1}$ a
pseudo-tree. In the third step, we add the new check node $C^*$ to
$\mathcal{G}_{i}$ and regenerate pseudo-tree or encoding stopping
sets in $\mathcal{G}_{i} \cup C^*$. Notice that the new parity
check equation $C^*$ in~(\ref{equ8}) does not incur extra cost to
compute variables $b_1$, $b_2$, $\ldots$, $b_m$ since these
variables have already been computed in those generalized parity
check equations in $\mathcal{G}_{i+1}$, as shown in~(\ref{equ6}).
Practically, we can compute these variables $b_1$, $b_2$,
$\ldots$, $b_m$ only once and store them. Later, we can apply the
stored values $b_1$, $b_2$, $\ldots$, $b_m$ to both
equation~(\ref{equ8}) and equation~(\ref{equ6}). Hence, the new
parity check equation $C^*$ only needs $m-1$ additional XOR
operations to compute the summation of~$b_1$, $b_2$, $\ldots$,
$b_m$. Since the cost of encoding the pseudo-tree
$\mathcal{G}_{i+1}$ is $(m-1) \cdot (\overline{k}-2)$ where
$\overline{k}$ is the average degree of the remaining $m-1$~check
nodes in~$\mathcal{G}_{i+1}$, the overall cost of encoding
$\mathcal{G}_{i+1}$ and the new parity check equation~$C^*$ is
$(m-1) \cdot (\overline{k}-1)$.

By continuing to find pseudo-tree or encoding stopping sets in
this way, we reach the stage where $\mathcal{G} \backslash
\{\mathcal{G}_1 \cup \mathcal{G}_2 \cup \ldots \cup
\mathcal{G}_i\} = \phi$ or $\mathcal{G} \backslash \{\mathcal{G}_1
\cup \mathcal{G}_2 \cup \ldots \cup \mathcal{G}_i\}$ does not
contain pseudo encoding stopping sets or encoding stopping sets.

By the above analysis, we decompose the Tanner graph~$\mathcal{G}$
into a sequence of $p$~subgraphs $\mathcal{G}_1$, $\mathcal{G}_2$,
$\ldots$, $\mathcal{G}_p$ where $\mathcal{G}_i$, $1 \leq i \leq
p$, is either a 1-fold-constraint encoding stopping set, a
2-fold-constraint encoding stopping set, or a pseudo-tree. If
$\mathcal{G}_i$ is a 1-fold-constraint or a 2-fold-constraint
encoding stopping set, we apply
Algorithm~\ref{encoding-algorithm2} or
Algorithm~\ref{encoding-algorithm4} to encode~$\mathcal{G}_i$ and
the resulting encoding complexity is less than~$2 \cdot M_i \cdot
(\overline{k_i} - 1)$ where $M_i$ denotes the number of
independent check nodes in~$\mathcal{G}_i$ and $\overline{k_i}$
denotes the average number of bit nodes contained in each check
node in~$\mathcal{G}_i$. If $\mathcal{G}_i$ is a pseudo-tree, we
apply Algorithm~\ref{encoding-algorithm1} to
encode~$\mathcal{G}_i$ and the corresponding encoding complexity
is less than~$M_i \cdot (\overline{k_i} - 1)$. The overall
computation complexity of encoding~$\mathcal{G}$ is linear on the
number of independent check nodes~$M$ in $\mathcal{G}$ and is
bounded by $\sum_{i=1}^p (2 \cdot M_i \cdot (\overline{k_i} - 1))
= 2 \cdot M \cdot (\overline{k} - 1)$ where $\overline{k}$ denotes
the average number of bits contained in each independent check
node of $\mathcal{G}$. This completes the proof. \hfill $\Box$

We summarize the algorithm of decomposing a Tanner graph with
maximum bit node degree~3 into pseudo-trees and encoding stopping
sets in Algorithm~\ref{decomposition} and the algorithm to encode
such LDPC codes in Algorithm~\ref{lineartimeencoding}.

\begin{algorithm}
\caption{Decompose a Tanner graph $\mathcal{G}$ with maximum bit
node degree 3 into 1-fold-constraint encoding stopping sets,
2-fold-constraint encoding stopping sets, and pseudo-trees.
\label{decomposition}}
\begin{algorithmic}
\STATE Find a pseudo encoding stopping set or an encoding stopping
set $\mathcal{G}_1$ from $\mathcal{G}$ using
Algorithm~\ref{encoding-algorithm9}. \STATE $\mathcal{G} =
\mathcal{G} \backslash \mathcal{G}_1$. \IF{$\mathcal{G}_1$ is a
pseudo encoding stopping set} \STATE Delete a check node in
$\mathcal{G}_1$. $\mathcal{G}_1$ becomes a pseudo-tree. \ENDIF
\STATE $i = 1$. \WHILE {there exists a pseudo encoding stopping
set or an encoding stopping set in $\mathcal{G}$}
       \STATE Find a pseudo encoding stopping set or an encoding stopping set $\mathcal{G}_{i+1}$ from
       $\mathcal{G}$ using Algorithm~\ref{encoding-algorithm9}. Assume $\mathcal{G}_{i+1}$ contain $m$ check nodes
       $C_1, C_2, \ldots, C_m$.
       \STATE $\mathcal{G} = \mathcal{G} \backslash \mathcal{G}_{i+1}$
       \IF {$\mathcal{G}_{i+1}$ is a pseudo encoding stopping set}
           \STATE $\mathcal{G}_{i+1} = \mathcal{G}_{i+1} \backslash C_m$.
           $\mathcal{G}_{i+1}$ becomes a pseudo-tree.
           \STATE Generate a new check node $C^* = C_1 \oplus C_2
           \oplus \ldots \oplus C_m$.
           \STATE Add $C^*$ to $\mathcal{G}_{i}$ and regenerate
           pseudo-trees and encoding stopping sets in $\mathcal{G}_{i} \cup C^*$.
       \ENDIF
       \STATE $i = i + 1$.
\ENDWHILE
    \STATE $\mathcal{G}_{i+1} = \mathcal{G}$.
    \STATE Output a sequence of subgraphs $\mathcal{G}_1$, $\mathcal{G}_2$, $\ldots$,
    $\mathcal{G}_p$ where $\mathcal{G}_i$, $1 \leq i \leq p$, is either a pseudo-tree
    or an encoding stopping set (1-fold-constraint or
    2-fold-constraint.)
\end{algorithmic}
\end{algorithm}

\begin{algorithm}
\caption{linear complexity encoding algorithm for LDPC codes with
maximum bit node degree 3 \label{lineartimeencoding}}
\begin{algorithmic}
\STATE {\bf Preprocessing (carry out only once):} \STATE Apply
Algorithm~\ref{decomposition} to decompose the Tanner graph
$\mathcal{G}$ of the code into $p$~subgraphs $\mathcal{G}_1$,
$\mathcal{G}_2$, $\ldots$, $\mathcal{G}_p$ where $\mathcal{G}_i$,
$i = 1, 2, \ldots, p$, is either a pseudo-tree or a
1-fold-constraint encoding stopping set or a 2-fold-constraint
encoding stopping set. \STATE {\bf Encoding:} \FOR {$i$ = 1 to
$p$}
              \STATE Compute the constants on the right-hand side
              of the generalized parity check equations of~$\mathcal{G}_i$
based on the already known bit values of~$\mathcal{G}_1$,
$\mathcal{G}_2$, $\ldots$, $\mathcal{G}_{i-1}$.
              \IF {$\mathcal{G}_i$ is a pseudo-tree}
           \STATE Encode $\mathcal{G}_i$ using Algorithm~\ref{encoding-algorithm1}.
       \ELSE
           \STATE Encode $\mathcal{G}_i$ using Algorithm~\ref{encoding-algorithm2}.
       \ENDIF

\ENDFOR
    \STATE Output the encoded codeword.

\end{algorithmic}
\end{algorithm}

Next, we extend the linear time encoding method described in
Theorem~\ref{thm:theorem3} to LDPC codes with arbitrary column
weight and row weight.
\begin{theorem}
\label{thm:theorem4} Any LDPC code~$\mathcal{C}$ with arbitrary
column weight distribution and row weight distribution can be
encoded in linear time, and the encoding complexity is less than
$4 \cdot M \cdot (\overline{k}-1)$ where $M$ is the number of
independent check nodes in $\mathcal{C}$ and $\overline{k}$ is the
average degree of check nodes.
\end{theorem}
\begin{figure}[htb]
\centerline{\epsfig{figure=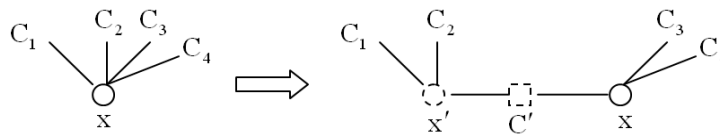,width=10cm,height=2cm}}
\caption{\label{encode9} Transform a bit node of degree~4 into two
bit nodes of degree~3 and an auxiliary check node.}
\end{figure}

\begin{figure}[htb]
\centerline{\epsfig{figure=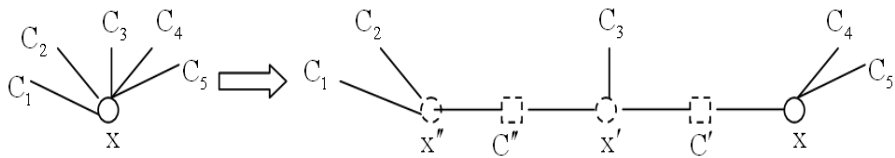,width=12cm,height=2.6cm}}
\caption{\label{encode10} Transform a bit node of degree~5 into
three bit nodes of degree~3 and two auxiliary check nodes.}
\end{figure}

\begin{figure}[htb]
\centerline{\epsfig{figure=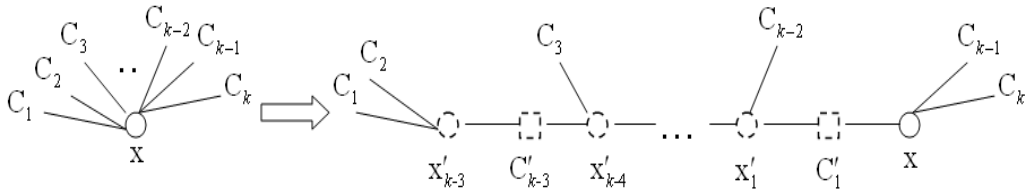,width=14cm,height=2.8cm}}
\caption{\label{encode11} Transform a bit node of degree~k into
$k-2$ bit nodes of degree~3 and $k-3$ auxiliary check node.}
\end{figure}
\emph{Proof}: We first show that an LDPC code with arbitrary
column weight distribution and row weight distribution can be
equivalently transformed into an LDPC code with maximum column
weight three. For example, Figure~\ref{encode9} on the left shows
a bit node~$x$ of degree~4. It can be split into two bit nodes~$x$
and~$x'$ of degree~3 and an auxiliary check node~$C'$, as shown on
the right in Figure~\ref{encode9}. The auxiliary check node~$C'$
is represented as~$x \oplus x' = 0$, which means $x$ is equivalent
to $x'$. Originally the bit node~$x$ connects to four check nodes
$C_1$, $C_2$, $C_3$, and $C_4$. After node splitting, $x'$
connects to $C_1$, $C_2$, and $x$ connects to $C_3$, $C_4$. Hence,
the Tanner graph on the left in Figure~\ref{encode9} is equivalent
to the Tanner graph on the right in Figure~\ref{encode9}.
Similarly, a bit node of degree~5 can be split into three bit
nodes~$x$, $x'$, and $x''$ and two auxiliary check nodes~$C'$
and~$C''$, as shown in Figure~\ref{encode10}. Generally, a bit
node of degree~$k$ can be equivalently transformed into $k-2$~bit
nodes of degree~3 and $k-3$~auxiliary check nodes, as shown in
Figure~\ref{encode11}. Assume an LDPC code~$\mathcal{C}$ contains
$M$~check nodes and $N$~bit nodes. The $M$ check nodes have
degrees $k_1$, $k_2$, $\ldots$, $k_M$, respectively. The $N$ bit
nodes have degrees $j_1$, $j_2$, $\ldots$, $j_N$, respectively.
Among the $N$ bit nodes in $\mathcal{C}$, there are $s$ bit nodes
whose degrees are greater than 3 and their degrees are $j_1$,
$j_2$, $\ldots$, $j_s$. This LDPC code can be equivalently
transformed into another LDPC code~$\mathcal{C'}$ with maximum
column weight 3. The new code $\mathcal{C'}$ has $M + \left
(\sum_{i=1}^s j_i - 3s\right )$ check nodes and $N + \left
(\sum_{i=1}^s j_i - 3s\right )$ bit nodes. By
Theorem~\ref{thm:theorem3}, the LDPC code $\mathcal{C'}$ can be
encoded in linear time and the encoding complexity is less than $2
\cdot M' \cdot (\overline{k'}-1)$. where $M'$ is the number of
independent check nodes in~$\mathcal{C'}$ and $\overline{k'}$ is
the average degree of independent check nodes in $\mathcal{C'}$.
Since there are $\sum_{i=1}^s j_i - 3s$ auxiliary check nodes in
$\mathcal{C'}$ that have degree~2, we derive that
\begin{eqnarray}
\label{equ29} \nonumber 2 \cdot M' \cdot (\overline{k'}-1) &=& 2
\cdot \left(M
\cdot (\overline{k} - 1) + \left(\sum_{i=1}^s j_i - 3s\right) \cdot (2-1)\right) \\
\nonumber &<& 2 \cdot \left(M \cdot (\overline{k} - 1) + \left(\sum_{i=1}^N j_i - N\right)\right) \\
\nonumber &<& 2 \cdot \left(M \cdot (\overline{k} - 1) + \left(\sum_{i=1}^M k_i - M\right)\right) \\
&=& 4 \cdot M \cdot (\overline{k} - 1)
\end{eqnarray} Therefore, the overall
computation cost of encoding $\mathcal{C}'$ is less than $4 \cdot
M \cdot (\overline{k} - 1)$. As the LDPC code $\mathcal{C}$ is
equivalent to the LDPC code $\mathcal{C}'$, the complexity of
encoding $\mathcal{C}$ is less than $4 \cdot M \cdot (\overline{k}
- 1)$. This completes the proof. \hfill $\Box$

Let's look at an example. The parity check matrix of a
(13,26)~LDPC code with column weight~3 is shown in~(\ref{equ9}).
Assume the values of the 13~information bits are 0, 1, 1, 1, 0, 1,
1, 0, 0, 1, 0, 1, and 1. We apply the proposed linear complexity
encoding method to encode this code.
\begin{equation}
\label{equ9} \mathbf{H}=\begin{array}{c} \left[
\begin{array}{cccccccccccccccccccccccccc}
                  1 & 0 & 0 & 0 & 0 & 0 & 0 & 0 & 1 & 0 & 0 & 0 & 1 & 0 & 0 & 0 & 0 & 0 & 0 & 0 & 1 & 1 & 1 & 0 & 0 & 0\\
                  0 & 0 & 0 & 1 & 0 & 0 & 0 & 1 & 0 & 0 & 0 & 1 & 0 & 0 & 0 & 1 & 0 & 1 & 0 & 0 & 0 & 0 & 1 & 0 & 0 & 0\\
                  0 & 0 & 1 & 0 & 0 & 0 & 0 & 0 & 0 & 0 & 1 & 0 & 0 & 0 & 1 & 0 & 0 & 0 & 0 & 1 & 0 & 0 & 0 & 0 & 1 & 1\\
                  0 & 0 & 0 & 0 & 0 & 0 & 1 & 1 & 0 & 1 & 0 & 0 & 0 & 1 & 0 & 0 & 0 & 0 & 0 & 0 & 1 & 0 & 0 & 1 & 0 & 0\\
                  0 & 0 & 0 & 0 & 0 & 1 & 0 & 0 & 0 & 0 & 0 & 0 & 0 & 0 & 1 & 0 & 0 & 0 & 1 & 1 & 0 & 0 & 0 & 0 & 1 & 1\\
                  0 & 0 & 0 & 0 & 1 & 0 & 1 & 0 & 0 & 0 & 0 & 1 & 1 & 0 & 0 & 0 & 1 & 0 & 0 & 0 & 0 & 0 & 1 & 0 & 0 & 0\\
                  0 & 0 & 0 & 0 & 1 & 0 & 0 & 0 & 1 & 0 & 0 & 0 & 1 & 1 & 0 & 1 & 0 & 1 & 0 & 0 & 0 & 0 & 0 & 0 & 0 & 0\\
                  0 & 0 & 1 & 0 & 0 & 1 & 0 & 0 & 0 & 0 & 1 & 0 & 0 & 0 & 0 & 0 & 0 & 0 & 1 & 0 & 0 & 0 & 0 & 1 & 1 & 1\\
                  0 & 1 & 0 & 1 & 0 & 0 & 1 & 0 & 0 & 0 & 0 & 0 & 0 & 0 & 0 & 1 & 0 & 0 & 0 & 0 & 0 & 1 & 0 & 1 & 0 & 0\\
                  0 & 1 & 0 & 0 & 0 & 0 & 0 & 0 & 1 & 1 & 0 & 1 & 0 & 1 & 0 & 0 & 0 & 1 & 0 & 0 & 0 & 0 & 0 & 0 & 0 & 0\\
                  0 & 0 & 1 & 0 & 0 & 1 & 0 & 0 & 0 & 0 & 1 & 0 & 0 & 0 & 1 & 0 & 0 & 0 & 1 & 1 & 0 & 0 & 0 & 0 & 0 & 0\\
                  1 & 1 & 0 & 0 & 0 & 0 & 0 & 1 & 0 & 0 & 0 & 0 & 0 & 0 & 0 & 0 & 1 & 0 & 0 & 0 & 1 & 0 & 0 & 0 & 0 & 0\\
                  1 & 0 & 0 & 1 & 1 & 0 & 0 & 0 & 0 & 1 & 0 & 0 & 0 & 0 & 0 & 0 & 1 & 0 & 0 & 0 & 0 & 1 & 0 & 0 & 0
                  & 0
                  \end{array} \right] \end{array}
\end{equation}

\begin{figure}[htb]
\centerline{\epsfig{figure=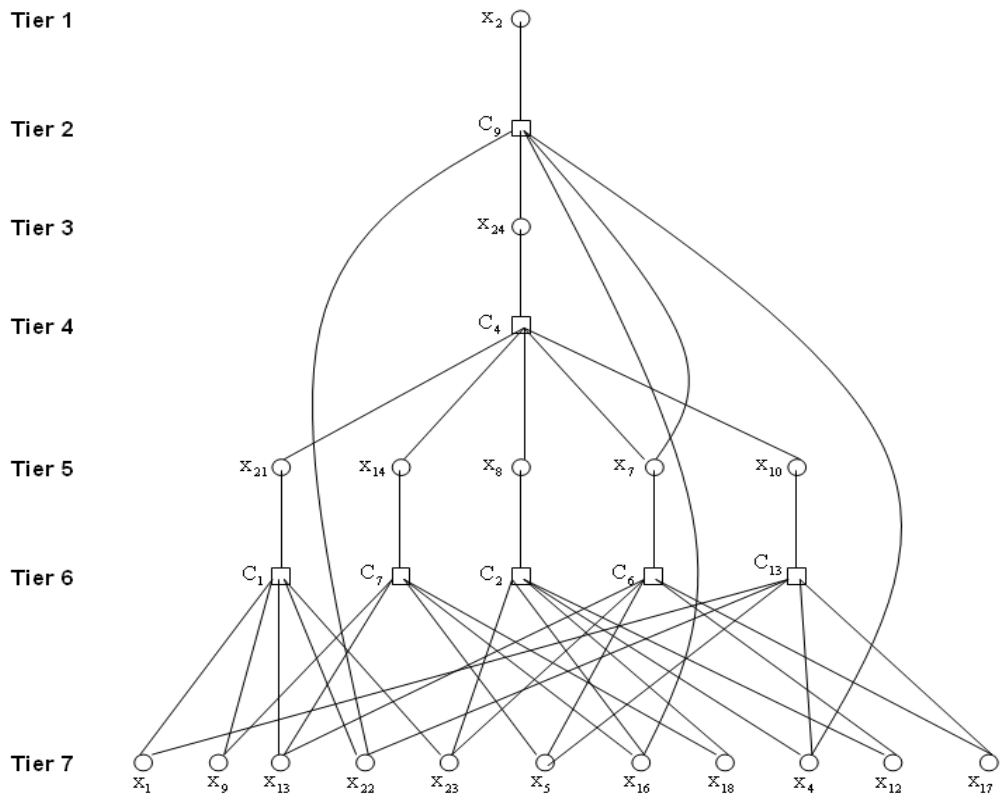,width=14cm}}
\caption{\label{encode12} A pseudo-tree built from the LDPC code
described in~(\ref{equ9}).}
\end{figure}

\begin{figure}[htb]
\centerline{\epsfig{figure=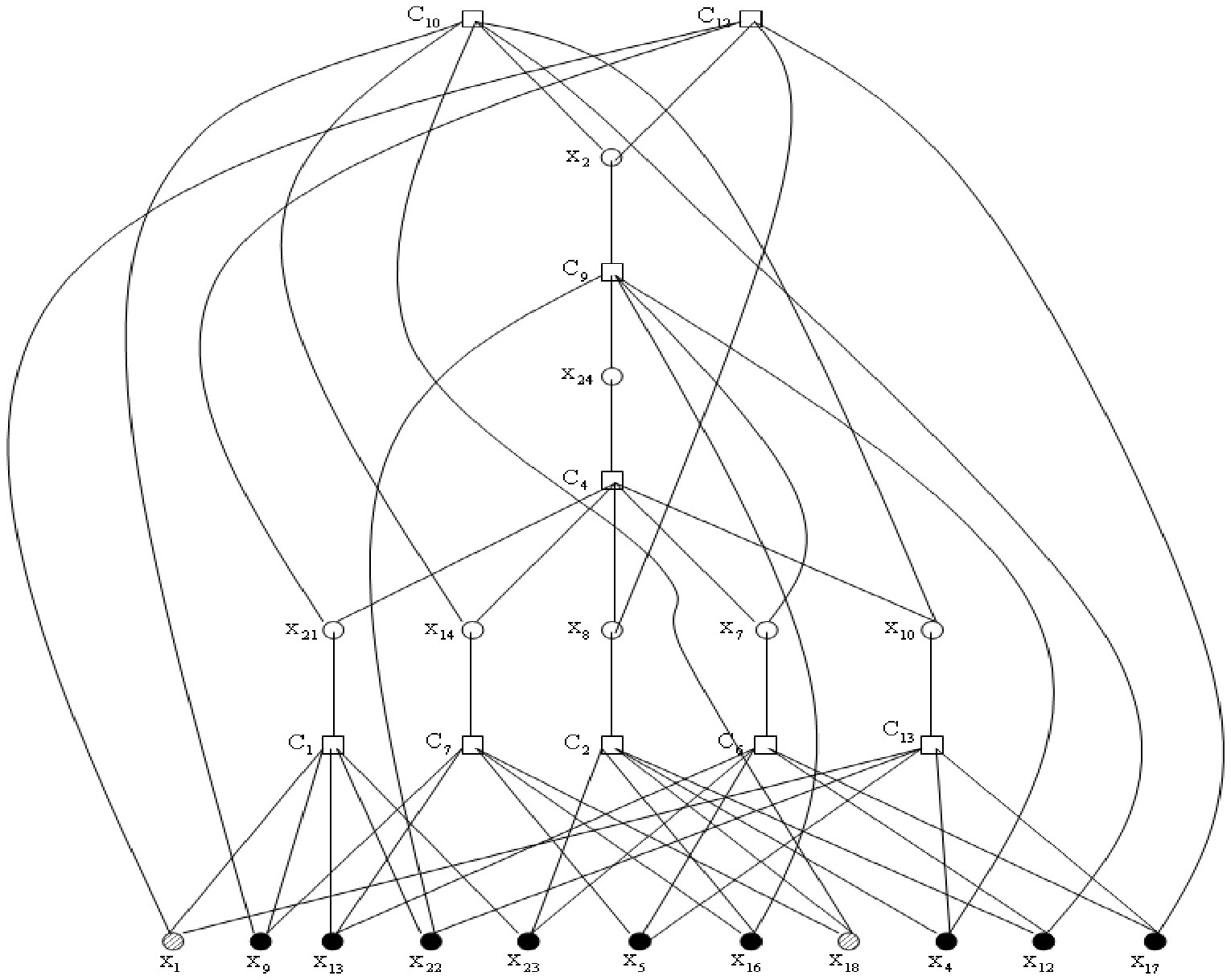,width=14cm}}
\caption{\label{encode13} An encoding stopping set developed from
the LDPC code described in~(\ref{equ9}).}
\end{figure}

{\bf Preprocessing.} We construct an encoding stopping set
from~(\ref{equ9}) using Algorithm~\ref{encoding-algorithm9}. We
start from an empty graph $\mathcal{S}$ and add check nodes and
their associated bit nodes to $\mathcal{S}$. Each time we add a
check node, we always pick the check node that contains the
smallest number of outsider nodes of $\mathcal{S}$. After adding 7
check nodes, the resulting graph is a pseudo-tree, as shown in
Figure~\ref{encode12}. When 9 check nodes are considered, we get
the 2-fold-constraint encoding stopping set~$\mathcal{E}_1$ shown
in Figure~\ref{encode13}. The bits $x_{9}, \,\, x_{13}, \,\,
x_{22}, \,\, x_{23}, \,\, x_{5}, \,\, x_{16}, \,\, x_{4}, \,\,
x_{12}, \,\, x_{17}$ are information bits. The two check nodes
$C_{10}$ and $C_{12}$ are key check nodes, and the two bit
nodes~$x_{1}$ and~$x_{18}$ are reevaluated bits.

\begin{figure}[htb]
\centerline{\epsfig{figure=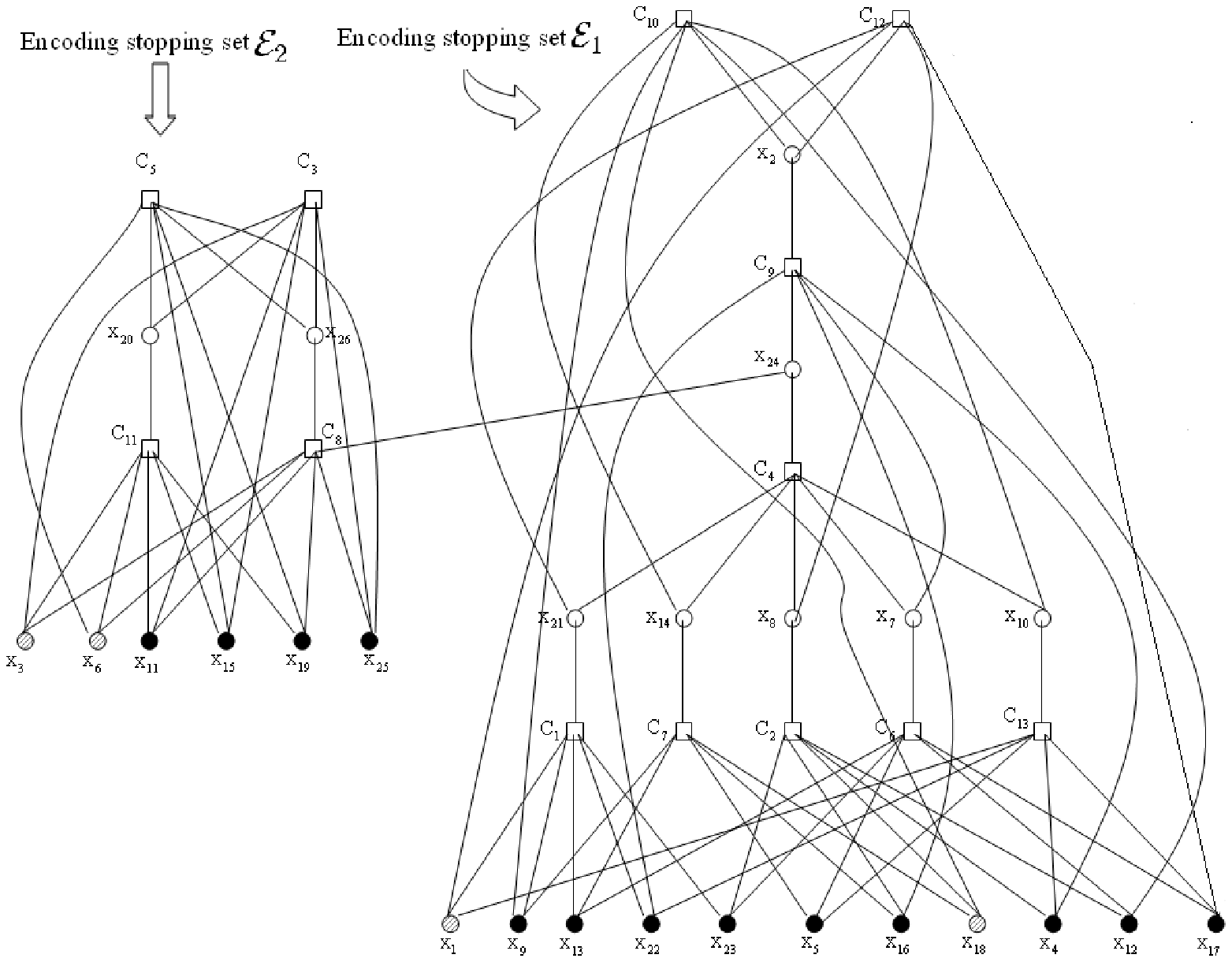,width=16cm}}
\caption{\label{encode14} Two encoding stopping sets developed
from the LDPC code described in~(\ref{equ9}).}
\end{figure}

After finding the encoding stopping set~$\mathcal{E}_1$, the
remaining Tanner graph of the code can be constructed to be a
2-fold-constraint encoding stopping set~$\mathcal{E}_2$, as shown
in Figure~\ref{encode14}. Therefore, the LDPC code can be
partitioned into two encoding stopping sets~$\mathcal{E}_1$
and~$\mathcal{E}_2$ that are shown in Figure~\ref{encode14}. The
bits $x_{11}, \,\, x_{15}, \,\, x_{19}, \,\, x_{25}$ in the
encoding stopping set~$\mathcal{E}_2$ are information bits. The
two check nodes $C_{3}$ and $C_{5}$ are key check nodes
of~$\mathcal{E}_2$, and the two bit nodes~$x_{3}$ and~$x_{6}$ are
reevaluated bits of~$\mathcal{E}_2$.

{\bf Encoding.}
\begin{itemize}
\item[{\bf Encode $\mathcal{E}_1$}:] \item[Step 1.] Fill the
values of the information bits, i.e., $[x_{9} \,\, x_{13} \,\,
x_{22} \,\, x_{23} \,\, x_{5} \,\, x_{16} \,\, x_{4} \,\, x_{12}
\,\, x_{17}]$ = [0 \,\, 1 \,\, 1 \,\, 1 \,\, 0 \,\, 1 \,\, 1 \,\,
0 \,\, 0]. Assign $x_1 = 0$ and $x_{18} = 0$. \item[Step 2.]
Encode the pseudo-tree shown in Figure~\ref{encode12}. Compute the
parity bits $x_{21}$, $x_{14}$, $x_{8}$, $x_{7}$, $x_{10}$,
$x_{24}$, $x_{2}$ as follows.
\begin{eqnarray}
\nonumber x_{21} &=& x_{1} \oplus x_{9} \oplus x_{13} \oplus
x_{22} \oplus x_{23} = 1
\\ \nonumber x_{14} &=& x_{9} \oplus x_{13} \oplus
x_{5} \oplus x_{16} \oplus x_{18} = 0 \\
\nonumber x_{8} &=& x_{23} \oplus x_{16} \oplus
x_{18} \oplus x_{4} \oplus x_{12} = 1 \\
\nonumber x_{7} &=& x_{13} \oplus x_{23} \oplus
x_{5} \oplus x_{12} \oplus x_{17} = 0 \\
\nonumber x_{10} &=& x_{1} \oplus x_{22} \oplus
x_{5} \oplus x_{4} \oplus x_{17} = 0 \\
\nonumber x_{24} &=& x_{21} \oplus x_{14} \oplus
x_{8} \oplus x_{7} \oplus x_{10} = 0 \\
\nonumber x_{2} &=& x_{22} \oplus x_{24} \oplus x_{16} \oplus
x_{7} \oplus x_{4} = 1
\end{eqnarray}
\item[Step 3.] Compute the values of the parity check
equations~$C_{10}$ and~$C_{12}$. $\widetilde{C_{10}} = x_{2}
\oplus x_{9} \oplus x_{10} \oplus x_{12} \oplus x_{14} \oplus
x_{18} = 1$, and $\widetilde{C_{12}} = x_{1} \oplus x_{2} \oplus
x_{8} \oplus x_{17} \oplus x_{21} = 1$. \item[Step 4.] Since
$\widetilde{C_{10}} = 1$  and $\widetilde{C_{12}} = 1$, the
correct values of~$x_{1}$ and $x_{18}$ are $x_{1} =
\widetilde{C_{10}} = 1$ and $x_{18} = \widetilde{C_{12}} = 1$.
\item[Step 5.] Recompute the parity bits~$x_{21}$, $x_{14}$,
$x_{8}$, and $x_{10}$ based on the new values of $x_{1}$ and
$x_{18}$. We derive that $x_{21} = 0$, $x_{14} = 1$, $x_{8} = 0$,
$x_{10} = 1$.
\item[{\bf Encode $\mathcal{E}_2$}:]
\item[Step 6.]
Fill the values of the information bits, i.e., $[x_{11} \,\,
x_{15} \,\, x_{19} \,\, x_{25}]$ = [1 \,\, 0 \,\, 1 \,\, 1].
Assign $x_3 = 0$ and $x_6 = 0$.
\item[Step 7.] Compute the parity
bits $x_{20}$, $x_{26}$ as follows.
\begin{eqnarray}
\nonumber x_{20} &=& x_{3} \oplus x_{6} \oplus x_{11} \oplus
x_{15} \oplus x_{19} = 0
\\ \nonumber x_{26} &=& x_{3} \oplus x_{6} \oplus
x_{11} \oplus x_{19} \oplus x_{25} \oplus x_{24} = 1
\end{eqnarray}

Notice that the value of the parity bit~$x_{26}$ is based on the
value of the bit~$x_{24}$ in~$\mathcal{E}_1$. \item[Step 8.]
Compute the values of the parity check equations~$C_{3}$
and~$C_{5}$. $\widetilde{C_{3}} = x_{3} \oplus x_{20} \oplus
x_{11} \oplus x_{15} \oplus x_{26} \oplus x_{25} = 1$, and
$\widetilde{C_{5}} = x_{6} \oplus x_{20} \oplus x_{15} \oplus
x_{19} \oplus x_{26} \oplus x_{25} = 1$. \item[Step 9.] Since
$\widetilde{C_{3}} = 1$ and $\widetilde{C_{5}} = 1$, the correct
values of~$x_{3}$ and $x_{6}$ are $x_{3} = \widetilde{C_{3}} = 1$
and $x_{6} = \widetilde{C_{5}} = 1$.
\end{itemize}
The encoded codeword is

$[x_{9} \,\, x_{13} \,\, x_{22} \,\, x_{23} \,\, x_{5} \,\, x_{16}
\,\, x_{4} \,\, x_{12} \,\, x_{17} \,\, x_{1} \,\, x_{18}\,\,
x_{21} \,\, x_{14} \,\, x_{8} \,\, x_{7} \,\, x_{10} \,\, x_{24}
\,\, x_{2} \,\, x_{11} \,\, x_{15} \,\, x_{19} \,\, x_{25} \,\,
x_{3} \,\, x_{6} \,\, x_{20} \,\, x_{26}] = [0 \,\, 1 \,\, 1 \,\,
1 \,\, 0 \,\, 1 \,\, 1 \,\, 0 \,\, 0 \,\, 1 \,\, 1\,\, 0 \,\, 1
\,\, 0 \,\, 0 \,\, 1 \,\, 0 \,\, 1 \,\, 1 \,\, 0 \,\, 1 \,\, 1
\,\, 1 \,\, 1 \,\, 0 \,\, 1]$

\section{Conclusion}
\label{sec:future}
This paper proposes a linear complexity
encoding method for general LDPC codes by analyzing and encoding
their Tanner graphs. We show that two particular types of Tanner
graphs--pseudo-trees and encoding stopping sets can be encoded in
linear time. Then, we prove that any Tanner graph can be
decomposed into pseudo-trees and encoding stopping sets. By
encoding the pseudo-trees and encoding stopping sets in a
sequential order, we achieve linear complexity encoding for
arbitrary LDPC codes. The proposed method can be applied to a wide
range of codes; it is not limited to LDPC codes. It is applicable
to both regular LDPC codes and irregular LDPC codes. It is also
good for both ``low density'' parity check nodes and
``medium-to-high density'' parity check nodes. In fact, the
proposed linear time encoding method is applicable to any type of
block codes. It removes the problem of high encoding complexity
for all long block codes that historically are commonly encoded by
matrix multiplication.

%\appendix

\section*{Appendix~A \\ Finding reevaluated bits $x_\gamma$ and $x_\delta$ in a 2-fold-constraint encoding stopping set}
\label{appA}
The details are described in
Algorithm~\ref{encoding-algorithm8}.
\begin{algorithm}
\caption{Finding reevaluated bits $x_\gamma$ and $x_\delta$ in a
2-fold-constraint encoding stopping set
\label{encoding-algorithm8}}
\begin{algorithmic}
 \STATE Represent the two key check
equations $C_\alpha$ and $C_\beta$ as functions of the information
bits only. Assume $C_\alpha$ is associated with $q$ information
bits $x_{\alpha_1}$, $x_{\alpha_2}$, $\ldots$, $x_{\alpha_q}$ and
$C_\beta$ is associated with $p$ information bits $x_{\beta_1}$,
$x_{\beta_2}$, $\ldots$, $x_{\beta_p}$.
\STATE $Flag \leftarrow
0$.
\FOR{$i$ = 1 to $q$}
    \IF {$x_{\alpha_i}$ is contained in
$C_\alpha$ but not in $C_\beta$}
        \STATE $Flag \leftarrow 1$.
        \STATE Choose the reevaluated bit $x_\gamma$ to be $x_\gamma = x_{\alpha_i}$.
        \STATE exit the for loop.
    \ENDIF
 \ENDFOR
\IF {$Flag = 1$}
      \STATE Choose the reevaluated bit $x_\delta$ to be $x_\delta
      = x_{\beta_p}$.
\ELSE
      \STATE Choose the reevaluated bit $x_\gamma$ to be $x_\gamma
      = x_{\alpha_q}$.
      \FOR{$i$ = 1 to $p$}
          \IF {$x_{\beta_i}$ is contained in
$C_\beta$ but not in $C_\alpha$}
              \STATE Choose the reevaluated bit $x_\delta$ to be $x_\delta = x_{\beta_i}$.
              \STATE exit the for loop.
          \ENDIF
      \ENDFOR
\ENDIF \STATE Output the two chosen reevaluated bits $x_\gamma$
and $x_\delta$.
\end{algorithmic}
\end{algorithm}

\bibliographystyle{IEEEbib}

\end{document}